\documentclass[11pt,twoside]{article}

\newcommand{\titleshort}{Hierarchical robust aggregation of sales forecasts in e-commerce}
\newcommand{\titlelong}{\textbf{Hierarchical robust aggregation of sales forecasts at aggregated levels in e-commerce,
based on exponential smoothing and Holt's linear trend method}}
\newcommand{\authorsshort}{Huard, Garnier, Stoltz}
\newcommand{\authorslong}{
Malo Huard\footnote{Universit{\'e} Paris-Saclay, CNRS, Laboratoire de math{\'e}matiques d'Orsay, 91405 Orsay, France; \email{malo.huard@math.u-psud.fr} and \email{gilles.stoltz@math.u-psud.fr}}\textsuperscript{\,\,\,,\!\!\!}
\footnote{Milvue, Paris Biotech Sant{\'e}, 24 Rue du Faubourg Saint-Jacques, 75014 Paris}~~---
R{\'e}my Garnier\footnote{Cdiscount, 120--126 Quai de Bacalan, 33000 Bordeaux; \email{remy.garnier@ext.cdiscount.com}}\textsuperscript{\,\,\,,\!\!\!}
\footnote{Universit{\'e} Paris Seine, Laboratoire AGM, 95302 Cergy-Pontoise, France}~~---
Gilles Stoltz\textsuperscript{$*$,}\footnote{HEC Paris, Jouy-en-Josas, France; \email{stoltz@hec.fr}}
}

\newcommand{\storydate}{\today} 

\usepackage[utf8]{inputenc}

\usepackage{hyperref}
\usepackage{url}
\usepackage{fancyhdr}
\usepackage{amsmath,amssymb,amsthm}
\usepackage{natbib}
\usepackage[paperwidth=210mm,
                paperheight=297mm,
                top=2.5cm,
                textheight=247mm,
                footnotesep=2\baselineskip,
                left=2cm,
                right=2cm,
                marginparwidth=1.7cm,
                headsep=20pt,
                dvips=false]{geometry}
\usepackage{enumitem}
\usepackage{graphicx}
\usepackage{color}
\usepackage{algorithm}
\usepackage{algorithmic}
\usepackage{lmodern,textcomp}
\usepackage{cancel}

\newcommand{\email}[1]{\href{mailto:#1}{\texttt{#1}}}

\renewcommand{\leq}{\leqslant}
\renewcommand{\geq}{\geqslant}

\newcommand{\R}{\mathbb{R}}

\renewcommand{\hat}{\widehat}
\newcommand{\wh}{\widehat}
\newcommand{\wt}{\widetilde}

\newcommand{\cC}{\mathcal{C}}
\newcommand{\cX}{\mathcal{X}}
\newcommand{\cL}{\mathcal{L}}
\newcommand{\cH}{\mathcal{H}}
\newcommand{\rmse}{\mathrm{\textsc{rmse}}}
\newcommand{\mae}{\mathrm{\textsc{mae}}}
\newcommand{\mape}{\mathrm{\textsc{mape}}}
\newcommand{\sgn}{\mathrm{sgn}}
\renewcommand{\epsilon}{\varepsilon}
\newcommand{\sub}{\mathrm{\tiny sub}}

\newcommand{\transp}{\mbox{\tiny T}}

\usepackage{color}
\usepackage{xcolor}
\definecolor{Red}{rgb}{1.00, 0.00, 0.00}
\definecolor{green}{RGB}{0, 150, 0}

\newtheorem{remark}{Remark}

\newtheorem{exam}{Example}

\newcommand{\remar}{\noindent\emph{Remark.~~}}

\makeatletter
\renewcommand*{\@seccntformat}[1]{\csname the#1\endcsname.\quad}
\makeatother

\title{\titlelong}
\author{\authorslong}

\date{\vspace{1cm} \storydate}

\begin{document}
\setlist{itemsep=3pt,topsep=3pt,parsep=3pt,partopsep=3pt}
\renewcommand\labelitemi{--}

\pagestyle{fancy}
\renewcommand{\footrulewidth}{0.4pt}

\maketitle

\vspace*{3pt}\noindent\rule{\linewidth}{0.8pt}\vspace{5pt}
\thispagestyle{empty}
\renewcommand{\arraystretch}{1.2}

\begin{abstract}
We revisit the interest of classical statistical techniques for sales forecasting
like exponential smoothing and extensions thereof (as Holt's linear trend method).
We do so by considering ensemble forecasts, given by several instances of these classical techniques
tuned with different (sets of) parameters, and by forming convex combinations of the elements of
ensemble forecasts over time, in a robust and sequential manner. The machine-learning theory behind this is
called ``robust online aggregation'', or ``prediction with expert advice'',
or ``prediction of individual sequences'' (see \citealp{CBL}).
We apply this methodology to a hierarchical data set of
sales provided by the e-commerce company Cdiscount and output forecasts at the levels of
subsubfamilies, subfamilies and families of items sold, for various forecasting horizons (up to 6--week-ahead).
The performance achieved is better than what would be obtained by optimally tuning the classical techniques on
a train set and using their forecasts on the test set. The performance is also good from an intrinsic point
of view (in terms of mean absolute percentage of error).
While getting these better forecasts of sales at the levels of subsubfamilies, subfamilies and families
is interesting per se, we also suggest to use them as additional features
when forecasting demand at the item level.
\medskip

\end{abstract}

\section{Introduction and Literature Review}

Sales data in e-commerce are highly dynamic and volatile:
reactive methods are required (and these methods are often sophisticated).
We provide a detailed discussion of these newer methods in Section~\ref{sec:lit:gal};
they stem from the machine learning toolbox.
On the other hand, in retail merchandising, classical statistical techniques for sales forecasting
like exponential smoothing and extensions thereof (as Holt's linear trend method)
are effective and have been widely used since the 1950s
(see \citealp{Gardner85,Gardner06} and \citealp{Hyndman08}).
Other such classical techniques include autoregressive models like ARIMA
and its variants (\citealp{BJ94}). A review of the use of these classical techniques
may be found in the monograph by \citet{chatfield_time-series_2001},
and a recent application to the forecasting of intraday arrivals at a call
center was proposed by~\citet{MS-time-series}.

The aim of this article is to forecast sales in e-commerce based on
exponential smoothing and extensions thereof. By ``based on'', we mean
that two layers will be considered in our methodology:
the first layer is to build several instances of exponential smoothing
and Holt's linear trend method (tuned with different parameters).
They will be called elementary predictors. The forecasts of
these elementary predictors are then combined, prediction step after prediction step,
via a so-called aggregation algorithm (see \citealp{CBL} for an introduction
to the field of robust online aggregation). The aggregation algorithms considered output convex weights,
that evolve over time in a reactive way depending on performance,
and the aggregated forecasts are simply given by convex combinations of the forecasts
issued by the elementary predictors.

Since we are dealing with e-commerce data, the items considered
are grouped into a hierarchy (of subsubfamilies, subfamilies, and families of products).
We only forecast sales at these aggregated levels (not for individual items),
which, admittedly, is an easier forecasting task (see \citealp{MC84}).
We do so by aggregating the forecasts of elementary predictors separately at
each node of the hierarchy and by reconciling the thus obtained aggregated forecasts
through a projection. Cross-series information is thus shared
through the hierarchical constraints.
Our methodology is fully automated, scalable, and robust---three key requirements
stated by \citet{Seeger16}.

Sales forecasting at these aggregated levels may be considered interesting per
se, but we also see it as a way to obtain extra features for demand forecasting
at the item level; these extra features (sales forecasts for all items of the same
subsubfamily) can then be provided as an extra input to the sophisticated and
reactive machine-learning methods currently constructed (see Section~\ref{sec:lit:gal} for a more
detailed literature review).

\subsection{Presentation of the Problem of Hierarchical Forecasting and of the Data Set}
\label{sec:introhierarchy}

What follows is detailed in Sections~\ref{sec:aim} and~\ref{sec:descrdata}.
Our data was provided by the e-commerce company Cdiscount and
spans from July 2014 to December 2017---a period of 182 weeks.
We use July 2014 to December 2016 as a training period (containing $130$ weeks),
and January 2017 -- December 2017 (containing $52$ weeks) as a test period;
the test period thus features all major commercial events (sales, Black Friday and
Christmas shopping, etc.).
The data set features the daily sales of 620,749 items hierarchically ordered
in 3,004 subsubfamilies, 570 subfamilies and 53 families. We add up
daily sales to get weekly sales. Many time series of weekly sales thus created
are intermittent (but as will get clearer in the sequel, we do not apply any specific
trick or tool to deal with intermittent demand).

Our notion of a hierarchy means that we organize the subsubfamilies, subfamilies and families
into a tree $\Gamma$, whose root node consists of total sales.
The sales (numbers of units sold, or money value) achieved at a node $\gamma$ (i.e., for a given subsubfamily, subfamily or family)
during week $t$ are denoted by $s_{t,\gamma}$.
Summation constraints are considered: e.g., if $\gamma \in \Gamma$ is some (sub)family
and $\cC(\gamma)$ denotes the (sub)subfamilies that belong to it, we have
\[
s_{t,\gamma} = \sum_{c \in \cC(\gamma)} s_{t,c}\,.
\]
An arbitrary collection of forecasts $\wh{f}_{t+h,\gamma}$ of the sales at an horizon of $h$ weeks,
where $\gamma$ spans the tree $\Gamma$, may be transformed into a collection $\wt{f}_{t+h,\gamma}$
of such forecasts abiding by the summation constraints indicated by $\Gamma$
by a projection onto a suitable vector space. We further detail this in Section~\ref{sec:proj}.
Such a projection actually shares information between related subsubfamilies,
subfamilies and families.

\paragraph{Related literature on hierarchical forecasting.}
We provide hierarchical predictions but in a simple manner,
actually in the simplest possible manner: by independently
computing forecasts at each node of the hierarchy and by reconciling
them by a projection step. For a description of fancier approaches to
hierarchical forecasting, we refer to the specific literature review
provided in the introduction of \citet{BH20}.

\subsection{Robust Aggregation \\ (a.k.a.\ Prediction with Expert Advice, Prediction of Individual Sequences)}
\label{sec:hlmethodo}

The methodology discussed in this section is described in detail in Sections~\ref{sec:holt} and~\ref{sec:aggreg}.
It aims at providing node-by-node forecasts (series of forecasts for each given node $\gamma \in \Gamma$
of the hierarchy).

Our methodology relies on ensemble forecasts (Section~\ref{sec:holt}): several elementary predictors are considered, all of them but a few given
by instances of exponential smoothing or Holt's linear trend method, with different sets of parameters.
As the series of sales all exhibit some seasonality, but with different cycles depending on the considered node $\gamma$,
as some have a linear trend and some others do not, as some are highly regular while some others exhibit a more erratic behavior,
it is clear that no single instance of exponential smoothing or Holt's linear trend method can be simultaneously suited for all series.
This is why we consider several such instances ($J$ instances), which gives rise to a collection
\[
\wh{s}^{(j)}_{t+h,\gamma}, \qquad j \in \{1,\ldots,J\}\,,
\]
of elementary forecasts for the value $s_{t+h,\gamma}$.
A typical way to deal with this issue is to tune instead the parameters on a train set and use the thus-tuned method on the test set; i.e.,
to select one given elementary predictor among the $J$ ones considered. We show that typical methodology
is consistently inferior on our data set to aggregating (combining) the forecasts of all the elementary predictors,
as described below.

There are actually various techniques to aggregate forecasts via machine-learning or statistical methods.
Some of these aggregation techniques deal with stochastic data: the observations to be forecast are modeled by some stochastic process. On the contrary, other techniques work on deterministic data and come with theoretical guarantees of performance even when the observations cannot be modeled by a stochastic process. Examples of popular aggregation methods include Bayesian model averaging (see \citealp{BMA-tuto} for a tutorial and
\citealp{BMA-ens} for an application to ensemble forecasts) and random forests (introduced by \citealp{RF}), both of them being stochastic approaches,
as well as robust online aggregation, which is a deterministic approach.
We are interested in the latter approach, given the erratic nature of the series of sales in e-commerce (they are notoriously difficult to model).

Robust online aggregation is also known as prediction of individual sequences, or prediction with expert advice (see the monograph by \citealp{CBL} and
references therein, see also the numerous references provided in Section~\ref{sec:aggreg}).
This sequential aggregation technique, developed in the 1990s, provides a robust framework to make forecasts on a regular (e.g., weekly) basis. It does not rely on any specific assumption or need for stochastic modeling; it may handle any (bounded) time series, possibly extremely erratic.
At each time step, a weighted average of the forecasts of the elementary predictors is
issued, where the (convex) weights $w^{(1)}_{t+h,\gamma},\ldots,w^{(J)}_{t+h,\gamma}$
used are picked based on the past performance of the elementary predictors:
\[
\wh{f}_{t+h,\gamma} = \sum_{j=1}^J w^{(j)}_{t+h,\gamma} \, \wh{s}^{(j)}_{t+h,\gamma}\,.
\]
These weights thus change over time,
which guarantees that the aggregation algorithm may quickly adapt to changes in the environment, a key feature for e-commerce
that batch forecasting methods (the methods that use a train set) do not possess. In a nutshell, the
robust online aggregation algorithms considered are online and adaptive by nature, which is an advantage over
batch methods that are less often updated.

These robust online aggregation algorithms also come with strong theoretical guarantees of performance:
they almost achieve or outperform the performance of the best elementary predictor (and in some cases, the best constant convex
combination of elementary predictors). We note that the algorithms we relied on
are recent and effective aggregation algorithms---much more effective
than, e.g., the one (Vovk's ``Aggregation Algorithm'') considered by~\citet{OR-dynamic-pricing}
to learn demand characteristics while simultaneously pricing items.

\paragraph{Previous successful applications of robust aggregation in other fields.}
They are detailed in Section~\ref{sec:lit:gal}.

\subsection{What We Do and What We Don't}

\paragraph{What we don't do.}
Our data set did not include key features like the real-time evolution of the price of the items
nor their availability in stock.
We therefore do not consider sales forecasting in relationship with the prices offered,
which is a vast field of research; see \citet{OR-data-driven}, \citet{MSOM-demand} and
\citet{OR-demand-learning} for the use of price experiments as a demand learning tool,
as well as \citet{OR-dynamic-pricing} again (and the numerous references cited in these
three articles). Neither do we couple sales forecasting with anything else (\citealp{OR-time-series}
couples them with adaptive inventory policies).
Also, we rather use the terminology ``sales forecasting'' instead of ``demand forecasting''
as we are unable to tag null sales as potential lost sales.

\paragraph{What we do.} We provide a general methodology for the hierarchical forecasting
of time series (any time series: not necessarily sales), which is widely applicable
to any problem where univariate time series methods would be suited; e.g.,
the forecasting of intraday arrivals at a call center as proposed by~\citet{MS-time-series}.
We use modern and effective robust online aggregation algorithms to do so (more modern
algorithms than in \citealp{OR-dynamic-pricing}). Finally, we demonstrate the success of
our methodology on a real data set provided by the e-commerce company Cdiscount. We actually
started from the business practice---this data set---to build our methodology.

\subsection{Brief Summary of the Numerical Results Obtained}

The numerical results obtained are discussed in detail in
Section~\ref{sec:3}. We illustrate the good performance of our forecasting methodology
in two manners.

First, we provide a study of relative performance and show that the aggregation algorithms
considered consistently outperform the natural benchmark given by the best locally predictors on the train
set (i.e., what is achieved by selecting, for each node, the best elementary predictor on the
train set, and by using it on the test set), by about $5\%$. This observation holds
in mean absolute error [MAE] and in root mean square error [RMSE],
for various forecasting horizons (from 1--week-ahead to 6--week-ahead).
We note that the performance of the aggregation algorithms does not vary much
by the algorithm.

Second, we study the absolute (intrinsic) performance achieved, by reporting
mean absolute percentages of errors. Aggregation algorithms obtain a global
MAPE of about 20\% (again, this is valid for different forecasting horizons).
This MAPE can be broken down by the level: it equals about 30\% for subsubfamilies.
These values correspond to the consideration of aggregated levels; we recall that we do not
work at the item level.

Finally, we provide some graphical evolutions of convex weights picked
over time, for different families and for total sales. In general, these weights
change much over time, which illustrates the flexibility and reactivity
of the aggregation algorithms over time.

\subsection{Outline of the Article}

The article is organized as follows.
Section~\ref{sec:lit:gal} reviews the literature on sales and demand forecasting,
including the approaches specific to e-commerce.
Section~\ref{sec:2} presents the methodology followed
while Section~\ref{sec:3} discusses the results obtained on our data set.

More precisely, Section~\ref{sec:2} starts with a statement
of our setting of hierarchical prediction of sales (Section~\ref{sec:aim}).
It then describes the elementary predictors considered,
based on exponential smoothing or on Holt's linear trend method (Section~\ref{sec:holt}).
The aggregation methodology briefly hinted at above is described in details in
Section~\ref{sec:aggreg1} and three specific aggregation algorithms
are stated, and adapted where needed, in Section~\ref{sec:aggreg2}
(and a general trick to boost their performance is provided in Section~\ref{sec:aggreg3}).
However, Section~\ref{sec:aggreg} is only concerned with node-by-node aggregation
and this is why Section~\ref{sec:proj} explains how the node-by-node aggregation results
may be extended for the entire hierarchy of nodes.

Then, Section~\ref{sec:3} first provides a detailed description of the real data set considered
and of its division into a train set and a test set (Section~\ref{sec:descrdata}),
and discusses the performance of the elementary predictors at various forecasting horizons
(Section~\ref{sec:perf-elem}).
The main results consist of a tabulation of the performance
achieved by the three aggregation algorithms studied, in MAE and RMSE
(Section~\ref{sec:res:main}) and in MAPE (Section~\ref{sec:mape}).
Two complementary studies are finally provided:
on the distributions of errors (Section~\ref{sec:robustness})
and on the evolution of the weights put on each elementary predictor
by the aggregation algorithms (Section~\ref{sec:weights}).

\subsection{Additional Literature Review}
\label{sec:lit:gal}

We provide additional references on two topics:
on the applications of robust online aggregation
and on sales and demand forecasting.

\subsubsection{On the Applications of Robust Online Aggregation}
\label{sec:lit:aggreg}

As the methodology of robust online aggregation
hinted at in Section~\ref{sec:hlmethodo} does not rely on any specific assumption or need for stochastic modeling,
and is therefore extremely general, it was already successfully applied on different applications.
The R package Opera written by~\citet{Opera} is now a popular tool to use this methodology
and it is difficult to cite all applications already performed.
However, among them, we may cite the forecasting of air quality (\citealp{Oz}), of electricity load (\citealp{EDF,EDFBis,BH20}),
of exchange rates (\citealp{ExchRates}), of oil and gas production (\citealp{IFPEN}).

However, while the methodology is general, the application to each specific domain is still challenging:
some theoretical adaptations might be needed (in the present case, dealing with a hierarchy), and
more importantly, proper elementary predictors need to be designed. The ones used for
the forecasting of electricity load are actually quite fancy (see a specific discussion below, in Section~\ref{sec:lit}),
and the same can be said for air quality (complex PDE models with different data inputs, see \citealp{Oz})
and oil and gas production (complex numerical solvers were used to model the production fields, see \citealp{IFPEN}).
In the present article, we want to show that classical and simple time series methods like exponential
smoothing and Holt's linear trend method can be useful elementary predictors. Of course, more complex forecasting models
(using more side information) could be used as elementary predictors.

Also, in most references of this paragraph, aggregation algorithms outputting linear weights were considered
(e.g., ridge regression), while we restrict our attention to convex weights (to get safer predictions: within the range of
forecasts issued by elementary predictors).

\subsubsection{On Sales and Demand Forecasting}
\label{sec:lit}

So far, we only discussed general references on time-series predictions
(for exponential smoothing, Holt's linear trend method, ARIMA models)
and on ensemble methods, and in particular, on robust aggregation (also known as
prediction with expert advice or prediction of individual sequences, see Section~\ref{sec:hlmethodo}).
This is because our approach is designed to be general and independent of the
specific context of application.
However, we now provide a literature review focused on the goal of the present contribution,
namely, sales and demand forecasting, and even more precisely, sales and demand forecasting for e-commerce. \medskip

Demand forecasting tackles the prediction of the level of demand for a product or a service in the future.
This demand may not match the exact number of sales for a product for different reasons (stock shortage, change of prices).
Demand forecasts has various applications, among others: electric load forecasting (see \citealp{electric_load} for a survey,
see also the aforementioned contributions by \citealp{EDF,EDFBis,BH20}); urban water demand forecasts (see \citealp{water_load} for a survey);
and sales forecasting. General surveys on sales forecasting (not centered on e-commerce) were written by \citet{fashion_retail}
and \citet{supply_chain}. These applications differ on a number of criteria.
First, the demand variable may be continuous (case of electric load) or discrete (case of sales in retail business), with different aggregated times step (daily, weekly, monthly). In some cases, the demand variable may also be intermittent (see \citealp{intermittent} and~\citealp{Seeger16}
for examples and details). Second, the forecasting horizons differ between the considered applications,
from short-term prediction to longer-term horizons. The exact definition of short- and long-term may differ with applications, but a prediction horizon of more than two week is a long-term horizon for most applications.
The issue is that the generally best method for a given problem may differ for long-term and short-term predictions (see discussions by~\citealp{water_load}).
Third, the dimensions of the demand variables may differ.
In the simplest case, a unique value for a given time step is to be predicted; however,
in more complex cases, several values are to be predicted, for exemple,
levels of sales of multiple items at a given or at various time steps.
These multiple items may be organized in a hierarchy of products (as we do) or in related groups
(see~\citealp{chapados}).
This is why each of these applications presents some specific challenges to tackle.
We now detail two popular applications: electricity load forecasting and sales forecasting for e-commerce.

\paragraph{Electricity load forecasting.}
Traditional time series methods
(based on exponential smoothing or autoregressive models, and their extensions like Holt's linear trend method or ARIMA models)
have of course been extensively used and tailored to the needs of this application.
For instance, a lot of attention was put to add seasonality to this kind of models (see \citealp{exp_smoothing,taylor2010triple}
for recent examples).

Other modern machine statistical methods have been introduced to overcome the limitation of traditional times series methods.
We cite two of them. First, generalized additive models [GAMs], used in a autoregressive way (with past load values as features)
and with additional covariates (e.g., meteorological variables), are now a standard and efficient method to forecast
the electricity load, at least for short-term horizons; see \citet{PiGo11} and \citet{wijaya2015_gam}, as well as
their use by \citet{EDF} combined with robust aggregation.
Such GAM models rely on a discretization of the load into a sequence of values within a day by considering aggregated time steps, typically, half hours.
On the contrary, a second family of statistical models directly predicts load curves;
see \citet{Curve} and the discussions therein.
Of course, other methods, from the machine learning community, that may suffer from
a lack of interpretability, were also considered, like random forests: see \citet{Dudek}.

We may now provide a detailed comparison to the study by~\citet{BH20}, as promised
in Section~\ref{sec:introhierarchy}. The focus therein is the short-term
forecasting (one day ahead) of electricity load. Customers are grouped into a hierarchy
(created by clustering) so that the elementary predictors considered for
each cluster (based on sophisticated GAM models or on random forests) can be better adjusted.
Predictions are then formed cluster by cluster through robust aggregation algorithms
and reconciled through a projection step, exactly as in the present article.
Actually, the present article was initiated before and inspired the study by~\citet{BH20}.
More importantly, our focus here is to consider elementary predictors that are truly
elementary and general---and so are predictors based on exponential smoothing
and Holt's linear trend method, while GAM models or random forest are not.
The latter are powerful methods that are already efficient per se.
Finally, the hierarchies considered by~\citet{BH20} were small and limited,
while in the present article, we deal with a different scale (three layers and several thousands
of nodes and leaves).

\paragraph{Sales forecasting for e-commerce.}
This special case of demand forecasting comes with the following specific difficulties.
First, the number of items in e-commerce is generally large (much larger than in traditional retail);
these items are organized in a hierarchy of (subsub)families, as described in Section~\ref{sec:introhierarchy}.
Second, modern considerations in logistics and supply chain tend to limit supplies
and emphasizes \emph{just-on-time} resupply. This implies that e-commerce companies generally need
medium-term prediction for their sales, typically around 1-month (or 1-month-and-a-half) ahead.
However, most of the existing predictive models for supply chain were linear and were not able to deal with
the more erratic behaviour of real-world sales data in e-commerce.
Moreover, they were not able to exploit cross-product information.
This is why virtually all of the forecasting methods for sales in e-commerce
rely on sophisticated techniques stemming from the machine-learning community.
To name just a few, let us recall that a Bayesian modeling relying on a hierarchical state-space model was proposed
for sales data by~\citet{chapados} (and it allows to share information between products).
Neural network models have also been widely used, e.g., \citet{bandaraLSTM} used a recurrent neural network for e-commerce sales data.
Finally, Amazon developed a probabilistic neural network for demand forecast called DeepAR, described by \citealp{deepar}.
All these sophisticated methods are difficult to tune and maintain because
they rely on a large number of parameters; in contrast, our methodology is simple, computationally efficient,
and fully automated (once the elementary predictors are chosen).

\clearpage
\section{Setting}
\label{sec:2}

In this section we first describe the aim of the forecasting task
(Section~\ref{sec:aim}), the elementary predictors considered,
including Holt's linear trend predictors (Section~\ref{sec:holt}), and the aggregation methodology
followed. The latter first takes place node by node (Section~\ref{sec:aggreg}) and then
is extended to hold for the entire hierarchy of nodes (Section~\ref{sec:proj}).
The description of the node-by-node aggregation will be broken down into
a general presentation of the concept of aggregation (Section~\ref{sec:aggreg1}),
the statement of three specific aggregation algorithms considered in the sequel
(Section~\ref{sec:aggreg2}), and the description of the ``gradient trick'' (Section~\ref{sec:aggreg3}),
which is a general trick to boost the performance of aggregation algorithms.

\subsection{Aim: Hierarchical Prediction of Sales}
\label{sec:aim}

The products sold are grouped in a hierarchical way, given by a tree $\Gamma$;
nodes of the tree will be indexed by $\gamma$.
The root of $\Gamma$ gathers all products. The children of the root are called
families, and are further broken down into subfamilies, and then subsubfamilies.
The leaves of the tree correspond to the products.
A product corresponds to a unique subsubfamily, which itself corresponds to a unique subfamily,
which itself corresponds to a unique family.

We consider weekly sales, where weeks are indexed by $t \in \{1,2,\dots\}$.
We denote by $s_{t,\gamma}$ the sales achieved for family $\gamma$ during week $t$.
They can be measured in units or in total value.
The aim is to predict sales at all nodes of the hierarchy $\Gamma$, at a given horizon $h \geq 1$; that is, to
issue forecasts of the future quantities
\[
s_{t+h,\gamma}, \quad \gamma \in \Gamma\,.
\]
This aim was expressed to predict the sales during $n=1$ week, but we may possibly group $n \geq 2$ weeks,
with $n \leq h$, and forecast the quantities
\[
y_{t+h,\gamma} = \frac{1}{n} \sum_{\tau=t+h-n+1}^{t+h} s_{\tau,\gamma}, \quad \gamma \in \Gamma\,,
\]
which correspond to average sales over a period of $n$ weeks ending at the horizon of $h$ weeks.
Put differently, the goal is to forecast $h-n$--week-ahead a group of $n$ weeks (the group of $n$ weeks
starts at week $t+h-n+1$ after $h-n$ complete weeks have passed after the current week $t$).

The $y$ and the $s$ are equal in case $n = 1$, and this is why, with no loss of generality,
we only discuss below the forecast of the $y$. We defined the $y$ as averages for them to all share the
same order of magnitude, independently of the value of $n \geq 1$. Typical values for $n$ are
in $\{1,2,3,4\}$.

\paragraph{Summation constraints.}
The sales achieved at a given node are the sum of the sales achieved at its children nodes.
More formally, denoting by $\cC(\gamma)$ the children of a given node $\gamma \in \Gamma$,
we have, whenever $\cC(\gamma)$ is not the empty set:
\[
y_{t+h,\gamma} = \sum_{c \in \cC(\gamma)} y_{t+h,c}\,.
\]
It is thus natural to expect that the forecasts $\wh{y}$ of the $y$ satisfy
the same summation constraints: for all $\gamma \in \Gamma$
with non-empty set $\cC(\gamma)$ of children nodes,
\[
\wh{y}_{t+h,\gamma} = \sum_{c \in \cC(\gamma)} \wh{y}_{t+h,c}\,.
\]

\subsection{Elementary Predictors / Node by Node}
\label{sec:holt}

In this section, we fix a given node~$\gamma \in \Gamma$ and
describe the elementary forecasting methods considered. \medskip

We introduce three sets or families of elementary forecasting methods (or elementary predictors):
simple exponential smoothing, with an additive or a multiplicative treatment of seasonality, relying on a parameter $\alpha \in [0,1]$;
Holt's linear trend method, with an additive or a multiplicative treatment of seasonality,
relying on parameters $\alpha \in [0,1]$ and $\beta \in [0,1]$;
other elementary forecasts, provided by benchmarks.
Simple exponential smoothing and Holt's linear trend methods are popular
methods for demand forecasting in e--commerce (see \citealp{bandaraLSTM}).

Note that valid forecasts for the $y_{t+h,\gamma}$ quantities must rely
only on present and past sales, i.e., on sales $s_{\tau,\gamma}$ with $\tau \leq t$.
In particular, present and past average sales $y_{\tau,\gamma}$, with $\tau \leq t$, may
be used.

\paragraph{Other elementary forecasts.}
They consist of
\begin{itemize}
\item the sales achieved one year (52 weeks) ago, $\wh{y}_{t+h,\gamma} = y_{t+h-52,\gamma}$;
\item the sales currently achieved, $\wh{y}_{t+h,\gamma} = y_{t,\gamma}$;
\item the null sales, $\wh{y}_{t+h,\gamma} = 0$, given that a significant number of pairs of
subsubfamilies and weeks have no sales (most time series of sales are sparse, i.e., the demand of the
corresponding is intermittent; see the sparsity statistics
provided in Section~\ref{sec:descrdata}).
\end{itemize}
Here, and at all subsequent places,
the value $52$ weeks for a year could be replaced by $53$ weeks, which works equally well.
To alleviate notation, we did not set a parameter $T_{\mathrm{\scriptsize year}}$ for this value
but could have done so, of course.

\paragraph{Simple exponential smoothing with an additive treatment of seasonality.}
We use simple exponential smoothing to forecast the difference $d_{t+h,\gamma}$ between the quantity of interest,
$y_{t+h,\gamma}$, and its value one year ago, $y_{t+h-52,\gamma}$. This is a first (additive) way for
taking seasonality into account. Each instance of simple exponential smoothing is
parameterized by a number $\alpha \in [0,1]$.

More precisely, given the needed history, forecasts can only be issued after week $t_0$ (whose value is indicated below)
and are provided by
\[
\wh{d}_{t_0+h,\gamma} = d_{t_0,\gamma}
\qquad \mbox{and for $t \geq t_0+1$,} \qquad
\wh{d}_{t+h,\gamma} = \alpha \, d_{t,\gamma} + (1-\alpha) \, \wh{d}_{t-1+h,\gamma}\,;
\]
that is, $\wh{y}_{t_0+h,\gamma} = y_{t_0+h-52,\gamma} + (y_{t_0,\gamma} - y_{t_0-52,\gamma})$ and more generally, for $t \geq t_0$,
\[
\wh{y}_{t+h,\gamma} = y_{t+h-52,\gamma} + \sum_{j=0}^{t-t_0-1} \alpha (1-\alpha)^j \bigl( y_{t-j,\gamma} - y_{t-j-52,\gamma} \bigr)
+ (1-\alpha)^{t-t_0} \bigl( y_{t_0,\gamma} - y_{t_0-52,\gamma} \bigr)\,.
\]
The threshold $t_0$ is such that the $y$ with the smallest time index above, that is, $y_{t_0-52}$, is well defined;
it is defined as an average of $n$ weekly sales starting at time $t_0-52-n+1$, which must be at least $1$.
Thus, $t_0 = 52+n$.

\paragraph{Simple exponential smoothing with a multiplicative treatment of seasonality.}
A second (multiplicative) way for handling seasonality is to
replace the difference $d_{t+h,\gamma} = y_{t+h,\gamma} - y_{t+h-52,\gamma}$
by the ratio $y_{t+h,\gamma}/y_{t+h-52,\gamma}$.
We actually consider a variant of this ratio, given by
\[
z_{t+h,\gamma} = y_{t+h,\gamma}/r_{t+h-52,\gamma}\,,
\qquad \mbox{where} \qquad
r_{\tau,\gamma} = \frac{y_{\tau,\gamma}}{\displaystyle{\sum_{j=-26}^{25} y_{\tau+j,\gamma}}}
\]
denotes, for $\tau$ large enough, the ratio between the sales $y_{\tau,\gamma}$ for a given
week $\tau$ and yearly sales centered at this week.
Simple exponential smoothing is then used to forecast the $z$ quantities.

More precisely, given the needed history, forecasts can only be issued after a given week $t'_0$ (whose value is
indicated below) and are given by
\[
\wh{z}_{t'_0+h,\gamma} = z_{t'_0,\gamma}
\qquad \mbox{and, for $t \geq t'_0+1$,} \qquad
\wh{z}_{t+h,\gamma} = \alpha \, z_{t,\gamma} + (1-\alpha) \wh{z}_{t-1+h,\gamma}\,.
\]
(We skip the closed-form expressions that could be derived for the $\wh{z}_{t+h,\gamma}$.)
The forecasts of the quantities of interest are then provided, for $t \geq t'_0$, by
\[
\wh{y}_{t+h,\gamma} = r_{t+h-52,\gamma}\,\wh{z}_{t+h,\gamma}\,.
\]
The threshold $t'_0$ after which forecasts $\wh{y}_{t'_0+h}$ can be issued is such that
$\wh{z}_{t+h,\gamma} = z_{t'_0,\gamma}$ and $r_{t'_0+h-52,\gamma}$ are well defined.
It is necessary and sufficient to that end that $r_{t'_0-52,\gamma}$ be well defined.
The latter is an average of values $y_{\tau}$ starting at the index $t'_0-52-26$;
the starting value $y_{t'_0-52-26}$ is itself
an average of $n$ weekly sales starting at time $t'_0-52-26-n+1$, which must be at least $1$.
Thus, $t'_0 = 52+26+n$.

\paragraph{Holt's linear trend method with a multiplicative treatment of seasonality.}
We extend and generalize the approach followed in the previous paragraph
by allowing for a trend. Two parameters $\alpha \in [0,1]$ and $\beta \in [0,1]$
are set. The forecasting equations are, for $t \geq t'_0+2$ (where $t'_0$
was defined in the previous paragraph):
\begin{align*}
\mbox{[level]} \qquad\qquad  \phantom{e} & \ell_{t+h,\gamma} = \alpha \, z_{t,\gamma} + (1-\alpha) ( \ell_{t-1+h,\gamma} + b_{t-1+h,\gamma} ) \\
\mbox{[trend]} \qquad\qquad  \phantom{e} & b_{t+h,\gamma} = \beta ( \ell_{t+h,\gamma} - \ell_{t-1+h,\gamma} ) + (1-\beta) b_{t-1+h,\gamma}
\end{align*}
with an initialization consisting of
\[
\ell_{t'_0+1+h,\gamma} = z_{t'_0+1,\gamma}
\qquad \mbox{and} \qquad
b_{t'_0+1+h,\gamma} = z_{t'_0+1,\gamma} - z_{t'_0,\gamma}\,.
\]
The forecasts of the quantities of interest are then provided, for $t \geq t'_0+1$, by
\[
\wh{y}_{t+h,\gamma} = r_{t+h-52,\gamma}\,\bigl( \ell_{t+h,\gamma} + h \, b_{t+h,\gamma} \bigr)\,.
\]
\remar
The choice $\beta = 0$ with the initialization $b_{t'_0+1+h,\gamma} = 0$
(so that all $b$ values are null) corresponds to simple exponential smoothing.

\paragraph{Holt's linear trend method with an additive treatment of seasonality.}
We finally extend simple exponential smoothing with an additive treatment of seasonality
by also allowing for a trend; we use again the time $t_0 = 52+n$ defined therein.
The forecasting equations are, for $t \geq t_0+2$,
\begin{align*}
\mbox{[level]} \qquad\qquad  \phantom{e} & \ell_{t+h,\gamma} = \alpha (y_{t,\gamma} - y_{t-52,\gamma}) + (1-\alpha) ( \ell_{t-1+h,\gamma} + b_{t-1+h,\gamma} ) \\
\mbox{[trend]} \qquad\qquad  \phantom{e} & b_{t+h,\gamma} = \beta ( \ell_{t+h,\gamma} - \ell_{t-1+h,\gamma} ) + (1-\beta) b_{t-1+h,\gamma}
\end{align*}
with an initialization consisting of
\[
\ell_{t_0+1+h,\gamma} = y_{t_0+1,\gamma} - y_{t_0+1-52,\gamma}
\qquad \mbox{and} \qquad
b_{t_0+1+h,\gamma} = (y_{t_0+1,\gamma} - y_{t_0+1-52,\gamma}) - (y_{t_0,\gamma} - y_{t_0-52,\gamma})\,.
\]
The forecasts of the quantities of interest are then provided, for $t \geq t_0+1$, by
\[
\wh{y}_{t+h,\gamma} = y_{t+h-52,\gamma} + \bigl( \ell_{t+h,\gamma} + h \, b_{t+h,\gamma} \bigr)\,.
\]
\remar
The choice $\beta = 0$ with the initialization $b_{t_0+1+h,\gamma} = 0$
(so that all $b$ values are null) corresponds to simple exponential smoothing.

\subsection{Tuning Issue: Aggregating Rather Than Selecting Forecasts / Node by Node}
\label{sec:aggreg}

In this section, we describe the concept of
robust aggregation of predictors at a given node~$\gamma$.
The next section (Section~\ref{sec:proj}) will explain how to
extend this concept to predictions at all nodes of the hierarchy
considered.

\paragraph{Tuning issue.}
When only one set of forecasts (e.g., Holt's linear trend method
with a multiplicative treatment of seasonality) is considered, it suffices to tune
the two parameters $\alpha$ and $\beta$. This may typically be performed via cross-validation,
on a training set. This may be performed locally (the parameters $\alpha_\gamma$ and $\beta_\gamma$
picked depend on the node $\gamma$) or globally (the same parameters $\alpha$ and $\beta$ are used at all nodes).
However, in our case, several (sets of) elementary forecasts are available, which is more realistic. It may indeed be difficult
to determine beforehand whether seasonality should be addressed in an additive or a multiplicative way. Also, the simple
forecasts like the null sales may be particularly efficient for some subsubfamilies with rare sales.
This is why we rather resort to aggregation of elementary forecasts coming from various models
instead of selecting one particular forecasting method.
This methodology was developed in the machine learning community in the 1990s
and in the 2000s, see the monograph by \citet{CBL}. Its first application was
to construct portfolios to invest in the stock market (\citealp{Cover91})
and it has since then been successfully applied to a number of fields (see the end of Section~\ref{sec:hlmethodo}
for a detailed list).

To further describe the concept of aggregation of forecasts we discuss first the evaluation of
the forecasts issued.

\paragraph{Evaluating the quality of forecasts.}
We recall that sales $y$ may be evaluated in units or in total value (we will pick the latter
measure in our experiments). Two metrics are classically considered in logistics: the
mean absolute error [MAE] and the root mean square error [RMSE].

Consider a sequence $y_{1+h,\gamma},\ldots,y_{T+h,\gamma}$ of sales that were to be predicted
for a node $\gamma$, and assume that forecasts $\wh{y}_{1+h,\gamma},\ldots,\wh{y}_{T+h,\gamma}$ were issued. The MAE
and the RMSE of these forecasts are respectively defined by
\[
\mae = \frac{1}{T} \sum_{t=1}^T \bigl| y_{t+h,\gamma} - \wh{y}_{t+h,\gamma} \bigr|
\qquad \mbox{and} \qquad
\rmse = \sqrt{\frac{1}{T} \sum_{t=1}^T \bigl( y_{t+h,\gamma} - \wh{y}_{t+h,\gamma} \bigr)^2}\,.
\]

\subsubsection{Aggregation Methods: Principle and Guarantees}
\label{sec:aggreg1}

Since several elementary forecasting methods (possibly tuned with different sets of parameters),
say $J$ methods, we index their forecasts by a superscript $j \in \{1,\ldots,J\}$:
they provide the forecasts $\wh{y}^{(j)}_{t+h,\gamma}$.
At each prediction step, these elementary forecasts are combined in a convex way:
convex weights $w^{(1)}_{t+h,\gamma},\ldots,w^{(J)}_{t+h,\gamma}$ are picked,
i.e., non-negative numbers summing up to~$1$, and the aggregated forecast
\[
\wh{f}_{t+h,\gamma} = \sum_{j=1}^J w^{(j)}_{t+h,\gamma} \, \wh{y}^{(j)}_{t+h,\gamma}\,.
\]
Specific algorithms for picking these convex weights are described in Section~\ref{sec:aggreg2}.
Weights will be picked node by node.

The associated guarantees are typically of the following form:
at each node, the aggregated forecasts are at least almost as good as the best individual elementary forecasting
method, in MAE or in RMSE, while the aggregation algorithms do not know
in advance which elementary forecasting method is the most efficient.
In addition, no stochastic assumptions on the generating processes of the
sales or of the elementary forecasts are required.

More precisely, we denote by $[0,Y_\gamma]$ the range for the sales and forecasts of sales for node $\gamma$.
The MAE guarantees read: for all sequences of
sales $y_{t+h,\gamma} \in [0,Y_\gamma]$ and all sequences of elementary forecasts $\wh{y}^{(j)}_{t+h,\gamma} \in [0,Y_\gamma]$,
\begin{equation}
\label{eq:MAE-guar}
\frac{1}{T} \sum_{t=1}^T \bigl| y_{t+h,\gamma} - \wh{f}_{t+h,\gamma} \bigr|
\leq
\epsilon_{T,\gamma} +
\min_{j=1,\ldots,J} \,
\frac{1}{T} \sum_{t=1}^T \Bigl| y_{t+h,\gamma} - \wh{y}^{(j)}_{t+h,\gamma} \Bigr|\,, \qquad \mbox{where} \qquad \epsilon_{T,\gamma} \to 0\,.
\end{equation}
The bounds $\epsilon_{T,\gamma}$ only depend on $Y_\gamma$ and on $T$, they are uniform over
the sequences considered.

Similary, the RMSE guarantees read
\begin{equation}
\label{eq:RMSE-guar}
\sqrt{\frac{1}{T} \sum_{t=1}^T \bigl( y_{t+h,\gamma} - \wh{f}_{t+h,\gamma} \bigr)^2}
\leq
\epsilon'_{T,\gamma} +
\min_{j=1,\ldots,J} \,
\sqrt{\frac{1}{T} \sum_{t=1}^T \Bigl( y_{t+h,\gamma} - \wh{y}^{(j)}_{t+h,\gamma} \Bigr)^2}
\end{equation}
where the $\epsilon'_{T,\gamma}$ only depend on $Y_\gamma$ and on $T$ and satisfy $\epsilon'_{T,\gamma} \to 0$.

We now state the aggregation algorithms considered and hint at their associated guarantees, i.e.,
their associated values for the bounds $\epsilon_{T,\gamma}$ or $\epsilon'_{T,\gamma}$.

\subsubsection{Aggregation Methods: Three Examples}
\label{sec:aggreg2}

Three specific and popular aggregation algorithms are considered:
first, the polynomially weighted average forecaster with multiple learning rates [ML-Poly] and
the Prod forecaster with multiple learning rates [ML-Prod], both introduced
by \citet{MLP}; second, the Bernstein Online Aggregation [BOA] of \citet{BOA}.
Their statements in our context can be found in
Algorithms~\ref{algo:MLPoly}, \ref{algo:MLProd}, and~\ref{algo:BOA}.
The implementation of these algorithms depends on the guarantees~\eqref{eq:MAE-guar} or~\eqref{eq:RMSE-guar}
to be achieved. Indeed, as can be seen from their statements, they require a loss function: this should be the absolute
loss $\ell(y,f) = |y-f|$ in case the MAE guarantee~\eqref{eq:MAE-guar} is targeted, and the quadratic loss
$\ell(y,f) = (y-f)^2$ for the RMSE guarantee~\eqref{eq:RMSE-guar}.
Given our specific context, several adaptations with respect to the original statements of these
algorithms had to be performed, which are detailed below. We first provide some intuition
on what the various quantities maintained in the statements of the algorithms stand for,
and explain why we picked these algorithms.

\paragraph{Why these three algorithms? / What the various quantities maintained stand for.}
Both ML-Prod and BOA are variants of an alma matter aggregation algorithm called Hedge or
the exponentially weighted average [EWA] predictor, and introduced by \cite{Vovk90} and
\citet{LiWa94}. It relies on a learning rate $\eta > 0$
and picks weights (when adapted to our setting)
\[
w_{t+h,\gamma}^{(j)} = \frac{\exp \! \left( - \eta \displaystyle{\sum_{\tau=1}^t \ell\bigl(y_{\tau,\gamma},\wh{y}^{(j)}_{\tau,\gamma} \bigr)} \right)}{
\displaystyle{\sum_{k=1}^J \exp \! \left( - \eta \displaystyle{\sum_{\tau=1}^t \ell\bigl(y_{\tau,\gamma},\wh{y}^{(k)}_{\tau,\gamma} \bigr)} \right)}}
= \frac{\exp \! \left( \eta \displaystyle{\sum_{\tau=1}^t} \Biggl( \displaystyle{\sum_{i=1}^J} w^{(i)}_{t,\gamma} \, \ell\bigl(y_{t,\gamma},\wh{y}^{(i)}_{t,\gamma}\bigr) - \ell\bigl(y_{\tau,\gamma},\wh{y}^{(j)}_{\tau,\gamma} \bigr)\Biggr) \right)}{
\displaystyle{\sum_{k=1}^J} \exp \! \left( \eta \displaystyle{\sum_{\tau=1}^t}
\Biggl( \displaystyle{\sum_{i=1}^J} w^{(i)}_{t,\gamma} \, \ell\bigl(y_{t,\gamma},\wh{y}^{(i)}_{t,\gamma}\bigr) -
\ell\bigl(y_{\tau,\gamma},\wh{y}^{(k)}_{\tau,\gamma} \bigr) \Biggr)
\right)}
\]
ML-Prod is an adaptation of the second formulation of EWA on two main elements. First, the learning rate $\eta$
depends on each elementary predictor $k$ and is tuned over time: its value is given by
$f(S_{t,\gamma}^{(k)},S_{t,\gamma}^{(k)})$. Second, the exponential reweighting through the $\exp(-\eta x)$
function is replaced by a multiplicative update by $1-\eta x$, which is a first-order approximation of the exponent.
Similarly, BOA is an adaptation of the first formulation of EWA, where, in particular, prediction errors
\[
\ell\bigl(y_{\tau,\gamma},\wh{y}^{(j)}_{\tau,\gamma} \bigr)
\qquad \mbox{are replaced by} \qquad
\ell\bigl(y_{\tau,\gamma},\wh{y}^{(j)}_{\tau,\gamma} \bigr) \Bigl( 1 + \eta \ell\bigl(y_{\tau,\gamma},\wh{y}^{(j)}_{\tau,\gamma} \bigr) \Bigr)\,,
\]
which are only slightly larger quantities (as the learning rates are expected not to be too large).
ML-Prod and BOA were both designed based on EWA and carefully adapted to get better theoretical guarantees
and to not depend on any learning parameter (they are tuned automatically).
They are also known for exhibiting better performance in general than EWA
(see, e.g., discussions in the PhD thesis of \citealp{Gai15} and private feedback
collected from the users of the Opera package by \citealp{Opera}).

As for ML-Poly, it is an adaptation of the polynomially weighted average [PWA] predictor
(see \citealp{CeLu03}), which uses weights based on a polynomial reweighting scheme of the form
\[
w_{t+h,\gamma}^{(j)} =
\frac{\max \left\{0, \,\, \displaystyle{\sum_{\tau=1}^t} \Biggl( \displaystyle{\sum_{i=1}^J} w^{(i)}_{t,\gamma} \, \ell\bigl(y_{t,\gamma},\wh{y}^{(i)}_{t,\gamma}\bigr) - \ell\bigl(y_{\tau,\gamma},\wh{y}^{(j)}_{\tau,\gamma} \bigr)\Biggr) \right\}^{p-1}}{
\displaystyle{\sum_{k=1}^J}
\max \left\{0, \,\, \displaystyle{\sum_{\tau=1}^t} \Biggl( \displaystyle{\sum_{i=1}^J} w^{(i)}_{t,\gamma} \, \ell\bigl(y_{t,\gamma},\wh{y}^{(i)}_{t,\gamma}\bigr) - \ell\bigl(y_{\tau,\gamma},\wh{y}^{(k)}_{\tau,\gamma} \bigr)\Biggr) \right\}^{p-1}}\,,
\]
for some $p \geq 2$. ML-Poly corresponds to $p=2$ and will further reweight the nonnegative sums above (known as the
cumulative regret of each elementary predictor) by quantities denoted by $B^{(j)}_{t,\gamma} + S^{(j)}_{t,\gamma}$
in Algorithm~\ref{algo:MLPoly}.

The three algorithms discussed above are implemented ``from the book'' except for the needed adaptations described below.

\paragraph{Adaptations needed.}
First, the range of the prediction errors (i.e., of the loss functions) was assumed to be known in the original references,
while in our case, this range strongly depends on the numerous (subsub)families considered; there is no
reason for knowing the orders of magnitude of the sales, thus of the prediction errors, for each (subsub)family.
To cope for that, we maintain estimations $B_{t,\gamma}^{(j)}$ of the prediction errors (for BOA)
or squared excess prediction errors (for ML-Poly and ML-Prod) and use these estimates in lieu of the
known bounds of the original formulations of the algorithms.

Second, these algorithms were initially designed to forecast the next value of a time series, i.e.,
at time instance $t$, they issue forecasts of $y_{t+1}$. This corresponds, with our notation, to the case
$h = n = 1$. For other cases, we performed the adaptations relative to (i) the information available
at round $t$ when forecasting sales (ii) at an horizon $h$. For (i), we note that the grouped sales $y_{\tau,\gamma}$
involve averages over $n$ weeks, so that they are only defined for $\tau \geq n$;
for rounds $\tau \leq n$, the algorithms get no input and pick uniform aggregations
of the elementary forecasts. This is why time steps $t \in \{1,\ldots,n-1\}$ are handled separately.
For (ii), we use the value of the weights at round $t$ to aggregate the elementary forecasts
for the sales $y_{t+h,\gamma}$; this is in contrast with the original versions of the algorithms
where such a combination is performed to forecast the next element, not the next $h$--th element of the
time series.

Third, in the case of ML-Prod, the weight update
\[
W_{t,\gamma}^{(j)} = \Bigl( W_{t-1,\gamma}^{(j)} \Bigr)^{f(B_{t,\gamma}^{(j)},S_{t,\gamma}^{(j)})/f(B_{t-1,\gamma}^{(j)},S_{t-1,\gamma}^{(j)})}
\Bigl( 1 + f\bigl(B_{t,\gamma}^{(j)},S_{t,\gamma}^{(j)}\bigr) \, e_{t,\gamma}^{(j)} \Bigr)
\]
that may be read in Algorithm~\ref{algo:MLProd} slightly differs from the one that would have been
obtained ``from the book'', namely,
\[
W_{t,\gamma}^{(j)} = \biggl( W_{t-1,\gamma}^{(j)} \Bigl( 1 + f\bigl(B_{t-1,\gamma}^{(j)},S_{t-1,\gamma}^{(j)}\bigr) \, e_{t,\gamma}^{(j)} \Bigr) \biggr)^{f(B_{t,\gamma}^{(j)},S_{t,\gamma}^{(j)})/f(B_{t-1,\gamma}^{(j)},S_{t-1,\gamma}^{(j)})}\,;
\]
the former is a first-order approximation of the latter, and ensures that weights are well-defined: by definition of
all quantities maintained in the algorithm,
\[
\Bigl| f\bigl(B_{t,\gamma}^{(j)},S_{t,\gamma}^{(j)}\bigr) \, e_{t,\gamma}^{(j)} \Bigr|
\leq \frac{1}{2 B_{t,\gamma}^{(j)}} e_{t,\gamma}^{(j)} \leq \frac{1}{2 e_{t,\gamma}^{(j)}} e_{t,\gamma}^{(j)} \leq \frac{1}{2}\,,
\]
while no specific guarantee holds on $f\bigl(B_{t-1,\gamma}^{(j)},S_{t-1,\gamma}^{(j)}\bigr) \, e_{t,\gamma}^{(j)}$,
which could be smaller than $-1$ if $e_{t,\gamma}^{(j)}$ is a large negative number.

Without these adaptations, the three algorithms ensure theoretical guarantees~\eqref{eq:MAE-guar}
and~\eqref{eq:RMSE-guar} of respective orders $1/\sqrt{T}$ for $\epsilon_{T,\gamma}$ and $T^{-1/4}$ for $\epsilon'_{T,\gamma}$.
Such guarantees should still hold under the two adaptations performed
(estimated range and larger horizons $h \geq 2$).
The $T^{-1/4}$ rate for the RMSE is obtained
through an initial bound on the mean square errors of the form
\begin{equation}
\label{eq:fn}
\frac{1}{T} \sum_{t=1}^T \bigl( y_{t+h,\gamma} - \wh{f}_{t+h,\gamma} \bigr)^2
\leq
\bigl( \epsilon'_{T,\gamma} \bigr)^2 +
\min_{j=1,\ldots,J} \,
\frac{1}{T} \sum_{t=1}^T \Bigl( y_{t+h,\gamma} - \wh{y}^{(j)}_{t+h,\gamma} \Bigr)^2
\end{equation}
with $\bigl( \epsilon'_{T,\gamma} \bigr)^2$ of the order of $1/\sqrt{T}$, combined with
the inequality $\sqrt{a+b} \leq \sqrt{a} + \sqrt{b}$ for all non-negative numbers $a,b$.

\begin{algorithm}[p]
\caption{\label{algo:MLPoly} Polynomially weighted average forecaster with multiple learning rates
[ML-Poly], \newline \protect\phantom{\textbf{Algorithm~1}} plain version}
\begin{algorithmic}
\STATE \textbf{Parameters}
\STATE \quad Node $\gamma$, prediction horizon $h$, and week span $n$ with $1 \leq n \leq h-1$
\STATE \quad Loss function $\ell$ (absolute loss or quadratic loss) \\[5pt]

\STATE \textbf{Initialization}
\STATE \quad Set $R_{n-1,\gamma}^{(j)} = 0$, and
$B_{n-1,\gamma}^{(j)} = 0$, and $S_{n-1,\gamma}^{(j)} = 0$ for all $j \in \{1,\ldots,J\}$ \\[10pt]

\FOR{$t = 1,\ldots,n-1$}
\STATE Observe the elementary forecasts $\wh{y}^{(j)}_{t+h,\gamma}$, where $j \in \{1,\ldots,J\}$
\STATE Combine them uniformly, i.e., pick $w^{(j)}_{t+h,\gamma} = 1/J$ and form
$\displaystyle{\wh{f}_{t+h,\gamma} = \frac{1}{J} \sum_{j=1}^J \wh{y}^{(j)}_{t+h,\gamma}}$
\ENDFOR \\[10pt]

\FOR{$t = n, \, n+1, \,\ldots$}
\STATE Observe the sales $y_{t,\gamma}$
\renewcommand\algorithmicthen{}
\renewcommand\algorithmicdo{}
\FOR{$j \in \{1,\ldots,J\}$ \textbf{do} \hfill // For each elementary predictor $j$}
\renewcommand\algorithmicthen{\textbf{do}}
\renewcommand\algorithmicdo{\textbf{do}}
\STATE Set $\displaystyle{e_{t,\gamma}^{(j)} = \left( \sum_{k=1}^J w^{(k)}_{t,\gamma} \, \ell\bigl(y_{t,\gamma},\wh{y}^{(k)}_{t,\gamma}\bigr) \right) - \ell\bigl(y_{t,\gamma},\wh{y}^{(j)}_{t,\gamma}\bigr)}$ \hfill // Excess prediction error at $t$
\STATE Set $\displaystyle{R_{t,\gamma}^{(j)} = R_{t-1,\gamma}^{(j)} + e_{t,\gamma}^{(j)}}$ \hfill // Cumulated excess prediction error
\STATE Set $B_{t,\gamma}^{(j)} = \max \Bigl\{ B_{t-1,\gamma}^{(j)}, \,\, \bigl( e_{t,\gamma}^{(j)} \bigr)^2 \Bigr\}$
\hfill // Bound on squared excess errors
\STATE Set $S_{t,\gamma}^{(j)} = S_{t-1,\gamma}^{(j)} + \bigl( e_{t,\gamma}^{(j)} \bigr)^2$
\hfill // Sum of squared excess errors
\STATE Observe the elementary forecast $\wh{y}^{(j)}_{t+h,\gamma}$
\ENDFOR
\STATE Choose weights
\[
w^{(j)}_{t+h,\gamma} = \frac{\max\Bigl\{0,\,R^{(j)}_{t,\gamma}/ \bigl( B^{(j)}_{t,\gamma} + S^{(j)}_{t,\gamma} \bigr) \Bigr\}}{
\displaystyle{\sum_{k=1}^J \max\Bigl\{0,\,R^{(k)}_{t,\gamma}/ \bigl( B^{(k)}_{t,\gamma} + S^{(k)}_{t,\gamma} \bigr) \Bigr\}}}
\]
\STATE Form the aggregated forecast $\displaystyle{\wh{f}_{t+h,\gamma} = \sum_{j=1}^J w^{(j)}_{t+h,\gamma} \, \wh{y}^{(j)}_{t+h,\gamma}}$
\ENDFOR
\end{algorithmic}
\end{algorithm}

\begin{algorithm}
\caption{\label{algo:MLPoly-grad} ML-Poly, version with the gradient trick}
\begin{algorithmic}
\STATE Same as above, except for the line defining $e_{t,\gamma}^{(j)}$, which should be replaced by \\[5pt]
\STATE \qquad Set $e_{t,\gamma}^{(j)} = \psi\bigl(\wh{f}_{t,\gamma} - y_{t,\gamma}\bigr) \, \bigl( \wh{f}_{t,\gamma} - \wh{y}^{(j)}_{t,\gamma} \bigr)$ \\[5pt]
\STATE where $\psi(x) = 2x$ for the quadratic loss $\ell$ and $\psi(x) = \sgn(x)$ for the absolute loss $\ell$
\end{algorithmic}
\end{algorithm}

\begin{algorithm}
\caption{\label{algo:MLProd} Prod forecaster with multiple learning rates [ML-Prod], plain version}
\begin{algorithmic}
\STATE \textbf{Parameters}
\STATE \quad Node $\gamma$, prediction horizon $h$, and week span $n$ with $1 \leq n \leq h-1$
\STATE \quad Loss function $\ell$ (absolute loss or quadratic loss) \\[5pt]

\STATE \textbf{Notation}
\STATE \quad For $x,y > 0$, we define $\displaystyle{f(x,y) = \min \! \left\{ \frac{1}{2x}, \, \sqrt{\frac{\ln J}{x^2+y}} \right\}}$

\STATE \textbf{Initialization}
\STATE \quad Set $W_{n-1,\gamma}^{(j)} = 0$,
and $B_{n-1,\gamma}^{(j)} = 0$, and $S_{n-1,\gamma}^{(j)} = 0$ for all $j \in \{1,\ldots,J\}$ \\[10pt]

\FOR{$t = 1,\ldots,n-1$}
\STATE Observe the elementary forecasts $\wh{y}^{(j)}_{t+h,\gamma}$, where $j \in \{1,\ldots,J\}$
\STATE Combine them uniformly, i.e., pick $w^{(j)}_{t+h,\gamma} = 1/J$ and form
$\displaystyle{\wh{f}_{t+h,\gamma} = \frac{1}{J} \sum_{j=1}^J \wh{y}^{(j)}_{t+h,\gamma}}$
\ENDFOR \\[10pt]

\FOR{$t = n, \, n+1, \,\ldots$}
\STATE Observe the sales $y_{t,\gamma}$
\renewcommand\algorithmicthen{}
\renewcommand\algorithmicdo{}
\FOR{$j \in \{1,\ldots,J\}$ \textbf{do} \hfill // For each elementary predictor $j$}
\renewcommand\algorithmicthen{\textbf{do}}
\renewcommand\algorithmicdo{\textbf{do}}
\STATE Set $\displaystyle{e_{t,\gamma}^{(j)} = \left( \sum_{k=1}^J w^{(k)}_{t,\gamma} \, \ell\bigl(y_{t,\gamma},\wh{y}^{(k)}_{t,\gamma}\bigr) \right) - \ell\bigl(y_{t,\gamma},\wh{y}^{(j)}_{t,\gamma}\bigr)}$ \hfill // Excess prediction error at $t$
\STATE Set $B_{t,\gamma}^{(j)} = \max \Bigl\{ B_{t-1,\gamma}^{(j)}, \,\, \bigl| e_{t,\gamma}^{(j)} \bigr| \Bigr\}$ \hfill // Bound on excess errors
\STATE Set $S_{t,\gamma}^{(j)} = S_{t-1,\gamma}^{(j)} + \bigl( e_{t,\gamma}^{(j)} \bigr)^2$ \hfill // Cumulated excess prediction error \vspace{.2cm} \\
\STATE Set $\displaystyle{W_{t,\gamma}^{(j)} = \Bigl( W_{t-1,\gamma}^{(j)} \Bigr)^{f(B_{t,\gamma}^{(j)},S_{t,\gamma}^{(j)})/f(B_{t-1,\gamma}^{(j)},S_{t-1,\gamma}^{(j)})}
\Bigl( 1 + f\bigl(B_{t,\gamma}^{(j)},S_{t,\gamma}^{(j)}\bigr) \, e_{t,\gamma}^{(j)}} \Bigr)$
\STATE ~ \hfill // Multiplicative update of the weight maintained for predictor $j$ \vspace{.125cm} \\
\STATE Observe the elementary forecast $\wh{y}^{(j)}_{t+h,\gamma}$
\ENDFOR
\STATE Choose weights
\[
w^{(j)}_{t+h,\gamma} = \frac{f(B_{t,\gamma}^{(j)},S_{t,\gamma}^{(j)}) \, W^{(j)}_{t,\gamma}}{\displaystyle{\sum_{k=1}^J f(S_{t,\gamma}^{(k)},S_{t,\gamma}^{(k)}) \, W^{(k)}_{t,\gamma}}}
\]
\STATE Form the aggregated forecast $\displaystyle{\wh{f}_{t+h,\gamma} = \sum_{j=1}^J w^{(j)}_{t+h,\gamma} \, \wh{y}^{(j)}_{t+h,\gamma}}$
\ENDFOR
\end{algorithmic}
\end{algorithm}

\begin{algorithm}
\caption{\label{algo:MLProd-grad} ML-Prod, version with the gradient trick}
\begin{algorithmic}
\STATE Same as above, except for the line defining $e_{t,\gamma}^{(j)}$, which should be replaced by \\[5pt]
\STATE \qquad Set $e_{t,\gamma}^{(j)} = \psi\bigl(\wh{f}_{t,\gamma} - y_{t,\gamma}\bigr) \, \bigl( \wh{f}_{t,\gamma} - \wh{y}^{(j)}_{t,\gamma} \bigr)$ \\[5pt]
\STATE where $\psi(x) = 2x$ for the quadratic loss $\ell$ and $\psi(x) = \sgn(x)$ for the absolute loss $\ell$
\end{algorithmic}
\end{algorithm}

\begin{algorithm}
\caption{\label{algo:BOA} Bernstein Online Aggregation [BOA], plain version}
\begin{algorithmic}
\STATE \textbf{Parameters}
\STATE \quad Node $\gamma$, prediction horizon $h$, and week span $n$ with $1 \leq n \leq h-1$
\STATE \quad Loss function $\ell$ (absolute loss or quadratic loss) \\[5pt]

\STATE \textbf{Notation}
\STATE \quad For $x,y > 0$, we define $\displaystyle{f(x,y) = \min \! \left\{ \frac{1}{2x}, \, \sqrt{\frac{\ln J}{y}} \right\}}$

\STATE \textbf{Initialization}
\STATE \quad Set $L_{n-1,\gamma}^{(j)} = 0$, and $B_{n-1,\gamma}^{(j)} = 0$, and $S_{n-1,\gamma}^{(j)} = 0$,
and $\eta_{n-1,\gamma}^{(j)} = 0$
for all $j \in \{1,\ldots,J\}$ \\[10pt]

\FOR{$t = 1,\ldots,n-1$}
\STATE Observe the elementary forecasts $\wh{y}^{(j)}_{t+h,\gamma}$, where $j \in \{1,\ldots,J\}$
\STATE Combine them uniformly, i.e., pick $w^{(j)}_{t+h,\gamma} = 1/J$ and form
$\displaystyle{\wh{f}_{t+h,\gamma} = \frac{1}{J} \sum_{j=1}^J \wh{y}^{(j)}_{t+h,\gamma}}$
\ENDFOR \\[10pt]

\FOR{$t = n, \, n+1, \,\ldots$}
\STATE Observe the sales $y_{t,\gamma}$
\renewcommand\algorithmicthen{}
\renewcommand\algorithmicdo{}
\FOR{$j \in \{1,\ldots,J\}$ \textbf{do} \hfill // For each elementary predictor $j$}
\renewcommand\algorithmicthen{\textbf{do}}
\renewcommand\algorithmicdo{\textbf{do}}
\STATE Set $e_{t,\gamma}^{(j)} = \ell\bigl(y_{t,\gamma},\wh{y}^{(j)}_{t,\gamma}\bigr)$ \hfill // Prediction error at $t$
\STATE Set $L_{t,\gamma}^{(j)} = L_{t-1,\gamma}^{(j)} +  e_{t,\gamma}^{(j)} \, \Bigl( 1 + \eta^{(j)}_{t-1,\gamma}
e_{t,\gamma}^{(j)} \Bigr)$ \hfill // Cumulated (slightly enlarged) prediction errors
\STATE Set $B_{t,\gamma}^{(j)} = \max \Bigl\{ B_{t-1,\gamma}^{(j)}, \,\, e_{t,\gamma}^{(j)} \Bigr\}$
\hfill // Bound on prediction errors
\STATE Set $S_{t,\gamma}^{(j)} = S_{t-1,\gamma}^{(j)} + \bigl( e_{t,\gamma}^{(j)} \bigr)^2$
\hfill // Cumulative squared prediction errors
\STATE Set $\eta_{t,\gamma}^{(j)} = f\bigl(B_{t,\gamma}^{(j)},\,S_{t,\gamma}^{(j)}\bigr)$
\hfill // Weighting factor
\STATE Observe the elementary forecast $\wh{y}^{(j)}_{t+h,\gamma}$
\ENDFOR
\STATE Choose weights
\[
w^{(j)}_{t+h,\gamma} = \frac{\eta^{(j)}_{t,\gamma} \, \exp \bigl( - \eta^{(j)}_{t,\gamma} L^{(j)}_{t,\gamma} \bigr)}{\displaystyle{\sum_{k=1}^J
\eta^{(k)}_{t,\gamma} \, \exp \bigl( - \eta^{(k)}_{t,\gamma} L^{(k)}_{t,\gamma} \bigr)}}
\]
\STATE Form the aggregated forecast $\displaystyle{\wh{f}_{t+h,\gamma} = \sum_{j=1}^J w^{(j)}_{t+h,\gamma} \, \wh{y}^{(j)}_{t+h,\gamma}}$
\ENDFOR
\end{algorithmic}
\end{algorithm}

\begin{algorithm}
\caption{\label{algo:BOA-grad} BOA, version with the gradient trick}
\begin{algorithmic}
\STATE Same as above, except for the line defining $e_{t,\gamma}^{(j)}$, which should be replaced by \\[5pt]
\STATE \qquad Set $e_{t,\gamma}^{(j)} = \psi\bigl(\wh{f}_{t,\gamma} - y_{t,\gamma}\bigr) \, \wh{y}^{(j)}_{t,\gamma}$ \\[5pt]
\STATE where $\psi(x) = 2x$ for the quadratic loss $\ell$ and $\psi(x) = \sgn(x)$ for the absolute loss $\ell$
\end{algorithmic}
\end{algorithm}

\subsubsection{Comparison to the Best Convex Combination of Elementary Predictors \\ (= the Gradient Trick)}
\label{sec:aggreg3}

The guarantees~\eqref{eq:MAE-guar} and~\eqref{eq:RMSE-guar} can be strengthened, so that
the performance of the aggregation algorithm is almost as good as that of the
best constant convex combination of the elementary forecasts, i.e.,
the target
\[
\min_{(q_1,\ldots,q_J) \in \cX} \,
\frac{1}{T} \sum_{t=1}^T \left| y_{t+h,\gamma} - \sum_{j=1}^J q_j \, \wh{y}^{(j)}_{t+h,\gamma} \right|
\leq
\min_{j=1,\ldots,J} \,
\frac{1}{T} \sum_{t=1}^T \Bigl| y_{t+h,\gamma} - \wh{y}^{(j)}_{t+h,\gamma} \Bigr|
\]
is considered for MAE (and a similar target for RMSE), where $\cX$
denotes the set of all convex combinations, i.e.,
of all vectors $(q_1,\ldots,q_J)$ such that $q_j \geq 0$ for all $j$
and $q_1+\ldots+q_J=1$. Put differently, uniform bounds of the form
\[
\frac{1}{T} \sum_{t=1}^T \bigl| y_{t+h,\gamma} - \wh{f}_{t+h,\gamma} \bigr|
\leq
\epsilon_{T,\gamma} +
\min_{(q_1,\ldots,q_J) \in \cX} \,
\frac{1}{T} \sum_{t=1}^T \left| y_{t+h,\gamma} - \sum_{j=1}^J q_j \, \wh{y}^{(j)}_{t+h,\gamma} \right|,
\qquad \mbox{where} \qquad \epsilon_{T,\gamma} \to 0\,,
\]
and
\begin{multline*}
\sqrt{\frac{1}{T} \sum_{t=1}^T \bigl( y_{t+h,\gamma} - \wh{f}_{t+h,\gamma} \bigr)^2}
\leq
\epsilon'_{T,\gamma} +
\min_{(q_1,\ldots,q_J) \in \cX} \,
\sqrt{\frac{1}{T} \sum_{t=1}^T \left( y_{t+h,\gamma} - \sum_{j=1}^J q_j \, \wh{y}^{(j)}_{t+h,\gamma} \right)^{\!\! 2}}\,, \\
\mbox{where} \qquad \epsilon'_{T,\gamma} \to 0\,,
\end{multline*}
may be achieved, where the orders of magnitude of the $\epsilon_{T,\gamma}$ and $\epsilon'_{T,\gamma}$
are still $1/\sqrt{T}$ and $T^{-1/4}$.

To do so, the so-called ``gradient trick'' is applied (see, e.g., \citealp[Section~2.5]{CBL} and
references therein, in particular, \citealp{KiWa97} and \citealp{CB99}). It basically consists in replacing prediction errors by their gradients.
More precisely, the three algorithms stated above are modified as follows. In each statement, only the line defining
$e_{t,\gamma}^{(j)}$ based on the losses
$\ell\bigl(y_{t,\gamma},\wh{y}^{(k)}_{t,\gamma}\bigr)$ needs to be changed.
These losses are replaced by $\psi\bigl(\wh{f}_{t,\gamma} - y_{t,\gamma}\bigr) \, \wh{y}^{(j)}_{t,\gamma}$,
where $\psi : \R \to \R$ is defined as follows. For the quadratic loss $\ell(y,f) = (y-f)^2$,
we define $\psi(x)=2x$. For the absolute loss $\ell(y,f) = |y-f|$, we define $\psi(x) = \sgn(x)$, the sign of $x$,
that is,
\[
\sgn(x) = \left\{
\begin{array}{r}
+1 \mbox{ if } x>0 \\
0 \mbox{ if } x=0 \\
-1 \mbox{ if } x<0
\end{array}
\right.
\]
For the sake of clarity, the modified algorithms are stated below the original algorithms;
see Algorithms~\ref{algo:MLPoly-grad}, \ref{algo:MLProd-grad} and~\ref{algo:BOA-grad}.

\subsection{Providing Aggregated Forecasts for the Entire Hierarchy}
\label{sec:proj}

So far, we discussed the node-by-node prediction of sales, independently for each (subsub)family,
thus discarding for the time being the summation constraints indicated in
Section~\ref{sec:aim}. We now focus our attention on reconciling these independent
predictions.

\paragraph{Overall performance discarding the summation constraints.}
To that end, we first define the MAE and the RMSE of a family of sequences of forecasts over time
(similarly to what we did in Section~\ref{sec:aggreg} for a single sequence of forecasts over time).
Consider a family of sequences $y_{1+h,\gamma},\ldots,y_{T+h,\gamma}$ of sales that were to be predicted
for a hierarchy of nodes $\gamma \in \Gamma$,
and assume that families of sequences of forecasts $\wh{y}_{1+h,\gamma},\ldots,\wh{y}_{T+h,\gamma}$,
$\gamma \in \Gamma$, were issued.
The MAE and the RMSE of these families of sequences of forecasts are respectively defined by
\[
\mae = \frac{1}{T |\Gamma|} \sum_{t=1}^T \sum_{\gamma \in \Gamma} \bigl| y_{t+h,\gamma} - \wh{y}_{t+h,\gamma} \bigr|
\qquad \mbox{and} \qquad
\rmse = \sqrt{\frac{1}{T |\Gamma|} \sum_{t=1}^T \sum_{\gamma \in \Gamma} \bigl( y_{t+h,\gamma} - \wh{y}_{t+h,\gamma} \bigr)^2}\,,
\]
where $|\Gamma|$ denotes the cardinality of $\Gamma$.

When the guarantees~\eqref{eq:MAE-guar} and~\eqref{eq:RMSE-guar} hold for all $\gamma \in \Gamma$,
the overall performance achieved is almost as good as that of the best local elementary forecasting methods;
that is, by summing prediction errors along the hierarchy $\Gamma$, the following is guaranteed:
uniformly over sequences of sales and of elementary forecasts,
\begin{equation}
\label{eq:guar-MAE-joint}
\frac{1}{T |\Gamma|} \sum_{t=1}^T \sum_{\gamma \in \Gamma} \bigl| y_{t+h,\gamma} - \wh{f}_{t+h,\gamma} \bigr|
\leq
\epsilon_{T} +
\frac{1}{|\Gamma|} \sum_{\gamma \in \Gamma}
\min_{j=1,\ldots,J} \,
\frac{1}{T} \sum_{t=1}^T \Bigl| y_{t+h,\gamma} - \wh{y}^{(j)}_{t+h,\gamma} \Bigr|\,,
\qquad \mbox{where} \qquad \epsilon_{T} \to 0\,,
\end{equation}
and
\begin{equation}
\label{eq:guar-RMSE-joint}
\sqrt{\frac{1}{T |\Gamma|} \sum_{t=1}^T \sum_{\gamma \in \Gamma} \bigl( y_{t+h,\gamma} - \wh{f}_{t+h,\gamma} \bigr)^2}
\leq
\epsilon'_{T} +
\sqrt{\frac{1}{|\Gamma|} \sum_{\gamma \in \Gamma} \min_{j=1,\ldots,J} \, \frac{1}{T} \sum_{t=1}^T \Bigl( y_{t+h,\gamma} - \wh{y}^{(j)}_{t+h,\gamma} \Bigr)^2}\,,
\qquad \mbox{where} \qquad \epsilon'_{T} \to 0\,.
\end{equation}
(The bound in RMSE is obtained by first summing the initial bounds
described in~\eqref{eq:fn} and then taking square roots.)
The performance achieved by the best local elementary forecasting methods (the performance
reported in the right-hand sides above) will be called the oracle performance in the sequel.

\paragraph{Projections to abide by the summation constraints.}
Now, there is no reason for the aggregated forecasts
$\wh{f}_{t+h,\gamma}$ picked node by node as discussed in Section~\ref{sec:aggreg}
to abide by the summation constraints indicated in Section~\ref{sec:aim}.
This situation is similar to the one where a given elementary forecasting method
(e.g., Holt's linear trend method with a multiplicative treatment of seasonality)
is tuned node by node (e.g., by independent cross-validations), for the sake of efficiency:
possibly different parameters $\hat{\alpha}_\gamma,\,\hat{\beta}_\gamma$ are picked
for each node $\gamma$ and the elementary forecasts issued do not abide
by the summation constraints, in general.

A simple patch is however to project a vector of forecasts not abiding
by the summation constraints onto the vector space $\cH$ of those abiding by them;
formally, we define $\cH$ as the vector space of vectors $(f_\gamma)_{\gamma \in \Gamma}$
such that for all nodes $\gamma \in \Gamma$
with non-empty set $\cC(\gamma)$ of children nodes,
\[
f_{\gamma} = \sum_{c \in \cC(\gamma)} f_c\,.
\]
The projection may take place in Euclidean norm
or in absolute norm.
Let us denote by $\bigl(\wt{f}_{t+h,\gamma}\bigr)_{\gamma \in \Gamma}$ the projection
of $\bigl(\wh{f}_{t+h,\gamma}\bigr)_{\gamma \in \Gamma}$ onto $\cH$ in some norm
and let us review the theoretical guarantees, or lack thereof, associated with each norm.

\paragraph{Euclidean norm: theoretical guarantees.}
The theoretical guarantee that follows is already mentioned by~\citet{BH20}.
When the projection is in Euclidean norm, the Pythagorean theorem
ensures that
\[
\sqrt{\frac{1}{T |\Gamma|} \sum_{t=1}^T \sum_{\gamma \in \Gamma} \bigl( y_{t+h,\gamma} - \wt{f}_{t+h,\gamma} \bigr)^2}
\leq
\sqrt{\frac{1}{T |\Gamma|} \sum_{t=1}^T \sum_{\gamma \in \Gamma} \bigl( y_{t+h,\gamma} - \wh{f}_{t+h,\gamma} \bigr)^2}\,.
\]
Thus, whenever the guarantees~\eqref{eq:guar-MAE-joint} and~\eqref{eq:guar-RMSE-joint}
are satisfied for aggregated forecasts, they are also satisfied for their Euclidean projections.
The latter may only improve performance and ensure that the summation constraints
are satisfied, i.e., the forecasts issued are consistent with the hierarchy considered.

We implement the Euclidean projection $\Pi_{\cH}$ as follows.
We introduce the set $\cL(\Gamma)$ of leaves of $\Gamma$ and
a matrix $S$ indexed by $\Gamma \times \cL(\Gamma)$, where for all $\gamma \in \Gamma$ and
$\gamma' \in \cL(\Gamma)$,
\[
S_{\gamma,\gamma'} =
\left\{
\begin{array}{l}
1 \mbox{ if } \gamma = \gamma', \\
1 \mbox{ if } \gamma \mbox{ is the parent node of } \gamma', \\
0 \mbox{ otherwise}.
\end{array}
\right.
\]
The image of $S$ is exactly $\cH$. Since $S$ is injective and its image is $\cH$, it may
be shown that the Euclidean projection onto $\cH$ is given by the matrix
\[
\Pi_{\cH} = S (S^{\transp}S)^{-1} S^{\transp}\,.
\]

\paragraph{Absolute norm: no theoretical guarantee.}
The projection $\bigl(\wt{f}_{t+h,\gamma}\bigr)_{\gamma \in \Gamma}$
of $\bigl(\wh{f}_{t+h,\gamma}\bigr)_{\gamma \in \Gamma}$ onto $\cH$ in absolute norm is
defined as:
\[
\bigl(\wt{f}_{t+h,\gamma}\bigr)_{\gamma \in \Gamma}
\in \mathop{\mathrm{arg\,min}}_{(z_\gamma)_{\gamma \in \Gamma} \in \cH}
\frac{1}{T |\Gamma|} \sum_{t=1}^T \sum_{\gamma \in \Gamma} \bigl| z_{\gamma} - \wh{f}_{t+h,\gamma} \bigr|
\]
There are no theoretical guarantees on the performance of the projected forecasts,
as no Pythagorean-type theorem is able to relate
\[
\frac{1}{T |\Gamma|} \sum_{t=1}^T \sum_{\gamma \in \Gamma} \bigl| y_{t+h,\gamma} - \wt{f}_{t+h,\gamma} \bigr|
\qquad \mbox{to} \qquad
\frac{1}{T |\Gamma|} \sum_{t=1}^T \sum_{\gamma \in \Gamma} \bigl| y_{t+h,\gamma} - \wh{f}_{t+h,\gamma} \bigr|\,.
\]
Even worse, numerical results discussed in Section~\ref{sec:res:main} show that
the projection in absolute norm may even increase the prediction error.

\clearpage
\section{Numerical Results}
\label{sec:3}

We now apply the forecasting methodology described in the previous section to
our data set and more particularly, we consider the three algorithms
described therein (ML-Poly, ML-Prod and BOA) under the various
implementations possible: with a loss function given by the absolute loss or the quadratic loss,
with or without the gradient trick, with or without a Euclidean or absolute-norm
projection step after all local forecasts were issued (to meet the hierarchical constraints).

We compare the various implementations of these algorithms at different time horizons $(h,n)$.
We recall that $n$ denotes the number of weeks of sales considered in the forecasts and
$h$ the forecasting horizon, i.e., after the week considered,
there are $h-1$ weeks, and then starts the group of $n$ weeks to forecast;
the first week of this group is in $h$ weeks. Put differently, the group of weeks
to forecast is $h-1$--week-ahead.

\paragraph{Outline of the empirical study.}
We first provide a description of the real data set provided by
the company Cdiscount and how we divided it into a train set and a
test set (Section~\ref{sec:descrdata}). We then tabulate and graphically
illustrate the performance of the elementary predictors considered
(Section~\ref{sec:perf-elem}) depending on the cases $(h,n)$ considered;
the case $(h,n) = (7,1)$ is a challenging case, which is also representative of
a typical case from a business viewpoint.

We may then compare the three algorithms and their various implementations
on the case where $(h,n) = (7,1)$. We illustrate that the performance varies only slightly with
the algorithm picked and its specific implementation (loss function, gradient trick,
projection) and improves the locally best elementary predictors picked on the train set,
the natural benchmark, by a about $5\%$. This observation generalizes to all pairs $(h,n)$
considered (Section~\ref{sec:res:main}). So far, performance is studied only
in terms of MAE or RMSE. We then move (Section~\ref{sec:mape}) to an evaluation in terms
of mean absolute percentage of error [MAPE], to get a better grasp of the
forecasting performance (Section~\ref{sec:mape}). Again, aggregation methods improve by about $5\%$
the performance of natural benchmarks like the locally best elementary predictors picked on the train set,
achieving a MAPE of about $20\%$. This global MAPE is then broken down by the levels of the hierarchy,
and as expected, is larger for subsubfamilies (about $32\%$) than for subfamilies and families (about $22\%$ and
$18\%$) or for the total node (only about $12\%$).

Two complementary studies are finally provided.
As all results previously discussed were on average only,
we check that the better average performance obtained was so through
a shift of the distributions of errors towards zero (Section~\ref{sec:robustness}).
We also give an idea of how the weights put on each elementary predictor
evolve, on families: they are far from converging to anything and they show
that the aggregation methods are reactive to changes (Section~\ref{sec:weights}).

\subsection{Description of the Data Set}
\label{sec:descrdata}

Our data set is a real data set provided by the e-commerce company Cdiscount.
Our data spans from July 2014 to December 2017---a period of 182 weeks.
It features the daily sales of 620,749 products gathered in 3,004 subsubfamilies,
570 subfamilies and 53 families; that is, the cardinality of the
hierarchy $\Gamma$ is $3,\!004 + 570 + 53 + 1 = 3,\!628$ nodes,
including the leaves (subsubfamilies) and the root node (the total sales).
We added up daily sales to get weekly sales.

\begin{figure}[p]
\begin{center}
\includegraphics[width=\textwidth]{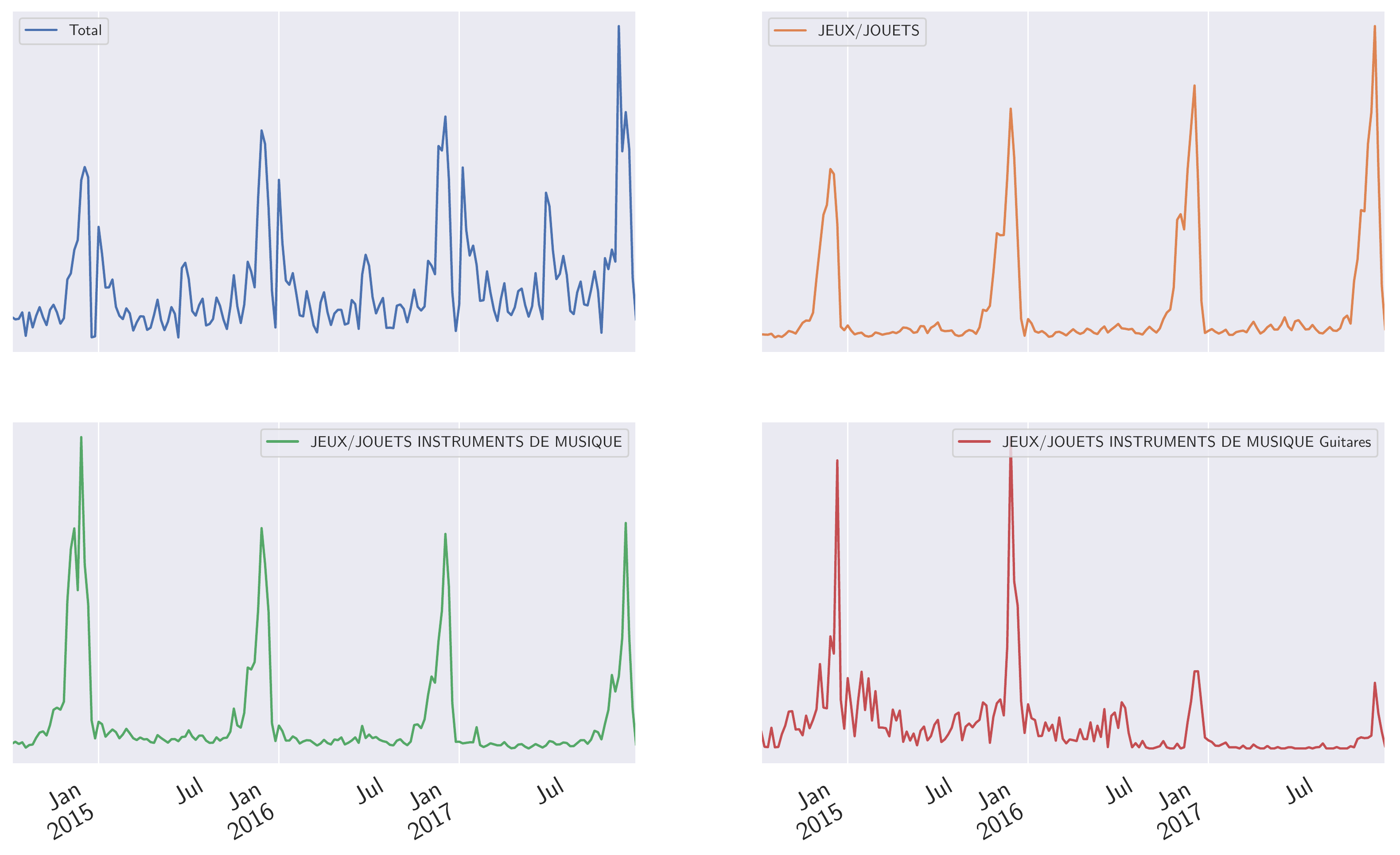} \\
\rule{0.75\linewidth}{0.8pt} \\
\ \\
\includegraphics[width=\textwidth]{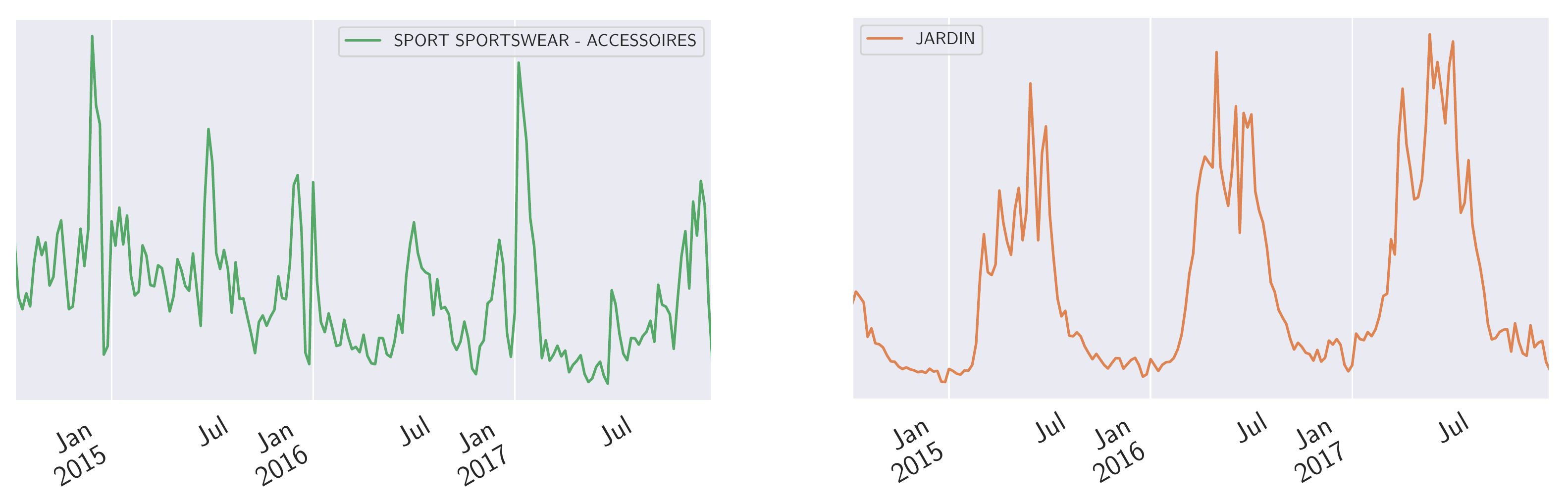}
\end{center}
\caption{\label{fig:descr} Weekly sales ($y$--\emph{values}) over time ($x$--\emph{axes}) at some nodes of the hierarchy.
The scales of the $y$--axes vary by graphs and are hidden for confidentiality
reasons. \smallskip \newline The \emph{top four graphs} correspond to a given path
in the hierarchy, corresponding to the subsubfamily of kids' guitars
({\color{red} red plot}), which is a part of the subfamily of
kids' music instruments ({\color{green} green plot}),
which itself belong to the family of toys ({\color{orange} orange plot}).
The evolution of the total sales at the root node is also provided ({\color{blue} blue plot}).
This path reads in French (see the legends on each graph):
Total {\textgreater} Jeux/Jouets {\textgreater} Instruments de musique {\textgreater} Guitares. \smallskip \newline
The \emph{bottom two graphs} feature the sales for the subsubfamily
({\color{green} green plot}, legend ``Sport - Sportswear Accessoires'') of sportswear accessories
and the family of garden products ({\color{orange} orange plot}, legend ``Jardin''), respectively.
}
\end{figure}

Figure~\ref{fig:descr} depicts some series of weekly sales:
the total sales (top left picture) and series associated with
two families, two subfamilies and one subsubfamily. These series
all exhibit some seasonality, but with different cycles. Some have a linear trend.
Some are highly regular, some others exhibit a more erratic behavior.
It is clear that no single elementary predictor of Section~\ref{sec:holt}
can be simultaneously suited for all series.

Table~\ref{tab:descr} provides some descriptive statistics (minimum and maximum, median
and means) on the weekly sales, by levels of the hierarchy of products.
This table also shows that many weekly-sales data points are null:
$45.3\%$ of the $3,\!004 \times 182$ weekly sales
for subsubfamilies, $48.3\%$ of the $570 \times 182$ weekly sales for
subfamilies, and even $38.1\%$ of the $53 \times 182$ weekly sales for
families. Part of these null values corresponds to intermittent demand, but it
turns out that some nodes of the hierarchy encounter null
sales during the entire period considered. More precisely, for 133 (out of $3,\!004$) subsubfamilies,
37 (out of 570) subfamilies, and 6 (out of 53) families, there are absolutely no sales
during the 182 weeks considered.
These high sparsity rates observed (on this data set and on other similar data sets of
e-commerce data) explain why the null elementary predictor defined in Section~\ref{sec:holt}
was considered.

\subsubsection*{Train set, test set}

We recall that our data spans from July 2014 to December 2017 (and features
$182$ weeks in total). We use July 2014 to December 2016 as a training period (containing $130$ weeks),
and January 2017 -- December 2017 (containing $52$ weeks) as a test period. The test period thus features all major commercial events (sales, Black Friday and
Christmas shopping, etc.). More precisely, after week $52+26+n \leq 80$ (given the values $n \leq 4$ considered below),
all elementary forecasting methods of Section~\ref{sec:holt} provide predictions and are aggregated via the algorithms described in
Section~\ref{sec:aggreg}, for the remaining part of the train period and also during the test period.
The performance obtained is however computed only on the test period, in MAE or RMSE, as explained at the beginning of Section~\ref{sec:aggreg}.

\vfill

\begin{table}[h]
\caption{\label{tab:descr} Some descriptive statistics on the weekly
sales (units: thousands of euros [k€]) for the 182 weeks considered, by
hierarchy level: subsubfamilies, subfamilies, families, and the root node (``Total'').
The numbers of nodes (``Count'') available for each level are recalled in the top
part of the table. Classical descriptive statistics (minimum and maximum, mean and median)
are provided in the middle part of the table. Specific descriptive statistics pertaining to the
sparsity of the sequences of weekly sales are given in the bottom part of the table:
the fraction of data points that are null (``Global sparsity rate'') among all data points
of this level, and counts of entire sequences of weekly sales that are null (``Null series: count'')
among the sequences of this level (there are ``Count'' of them).}
\begin{center}
\begin{tabular}{cccccc}
\hline
& Subsubfamilies & \phantom{e}Subfamilies\phantom{e} & \phantom{espe}Families\phantom{espe} & \phantom{e} & Total \\
\hline
Count & 3,004 & 570 & 53 & & 1 \\
\hline
Maximum & 5.6$\times 10^6$ & 9.4$\times 10^6$ & 17.5$\times 10^6$ & & 88.6$\times 10^6$\\
Mean & 10.1$\times 10^3$ & 53.4$\times 10^3$ & 574.6$\times 10^3$ & & 30.5$\times 10^6$ \\
Median & 18.4 & 7.1 & 246.8 & & 26.0$\times 10^6$ \\
Minimum & 0 & 0 & 0 & & 18.6$\times 10^3$ \\
\hline
Global sparsity rate & 45.3\% & 48.3\% & 38.1\% & & 0\% \\
Null series: count & 133 & 37 & 6 & & 0 \\
\hline
\end{tabular}
\end{center}
\end{table}

\clearpage

\subsection{Performance of the Elementary Forecasting Methods}
\label{sec:perf-elem}

\begin{figure}[p]
\begin{center}
\includegraphics[width=0.83\textwidth]{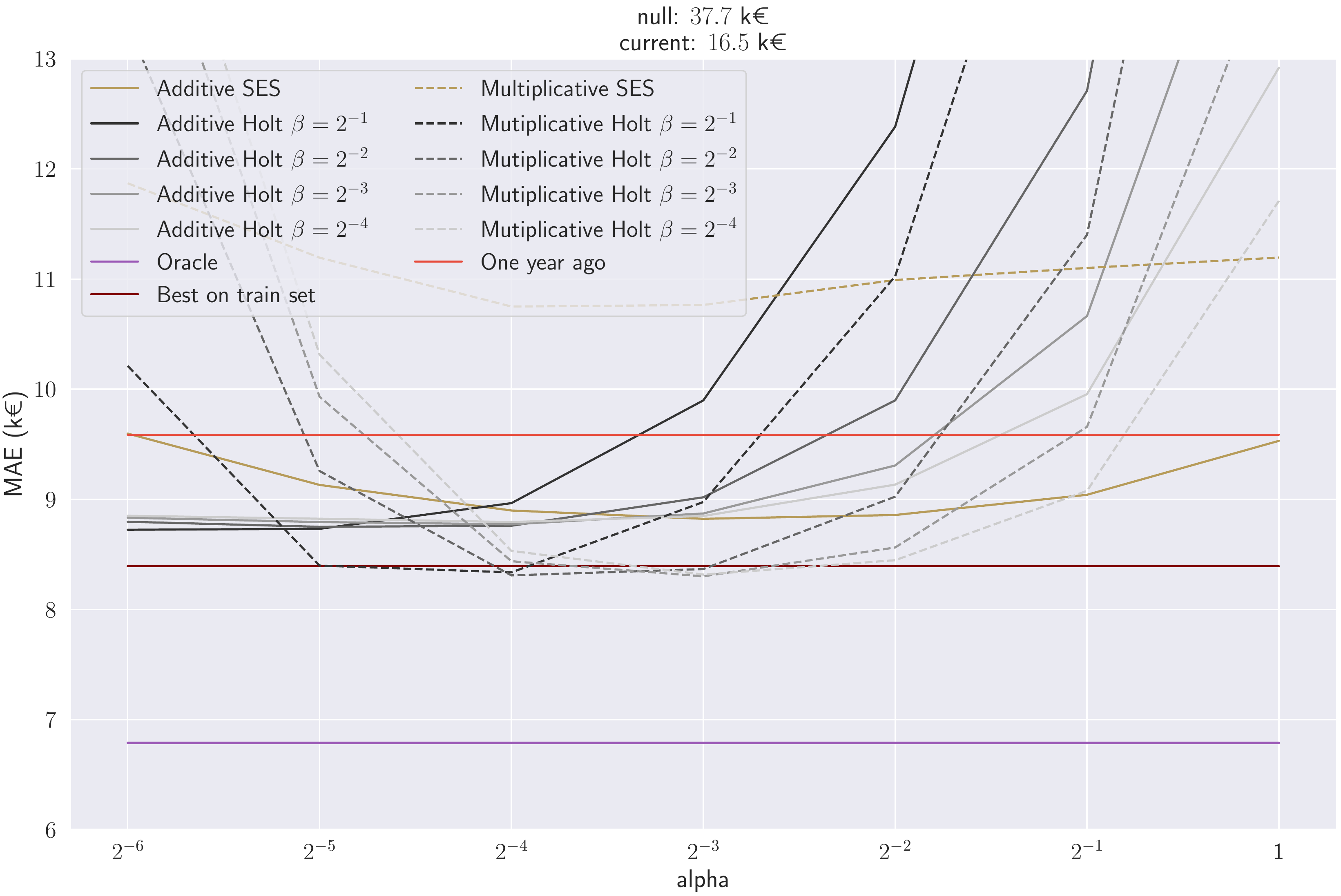} \\
\ \\
\includegraphics[width=0.83\textwidth]{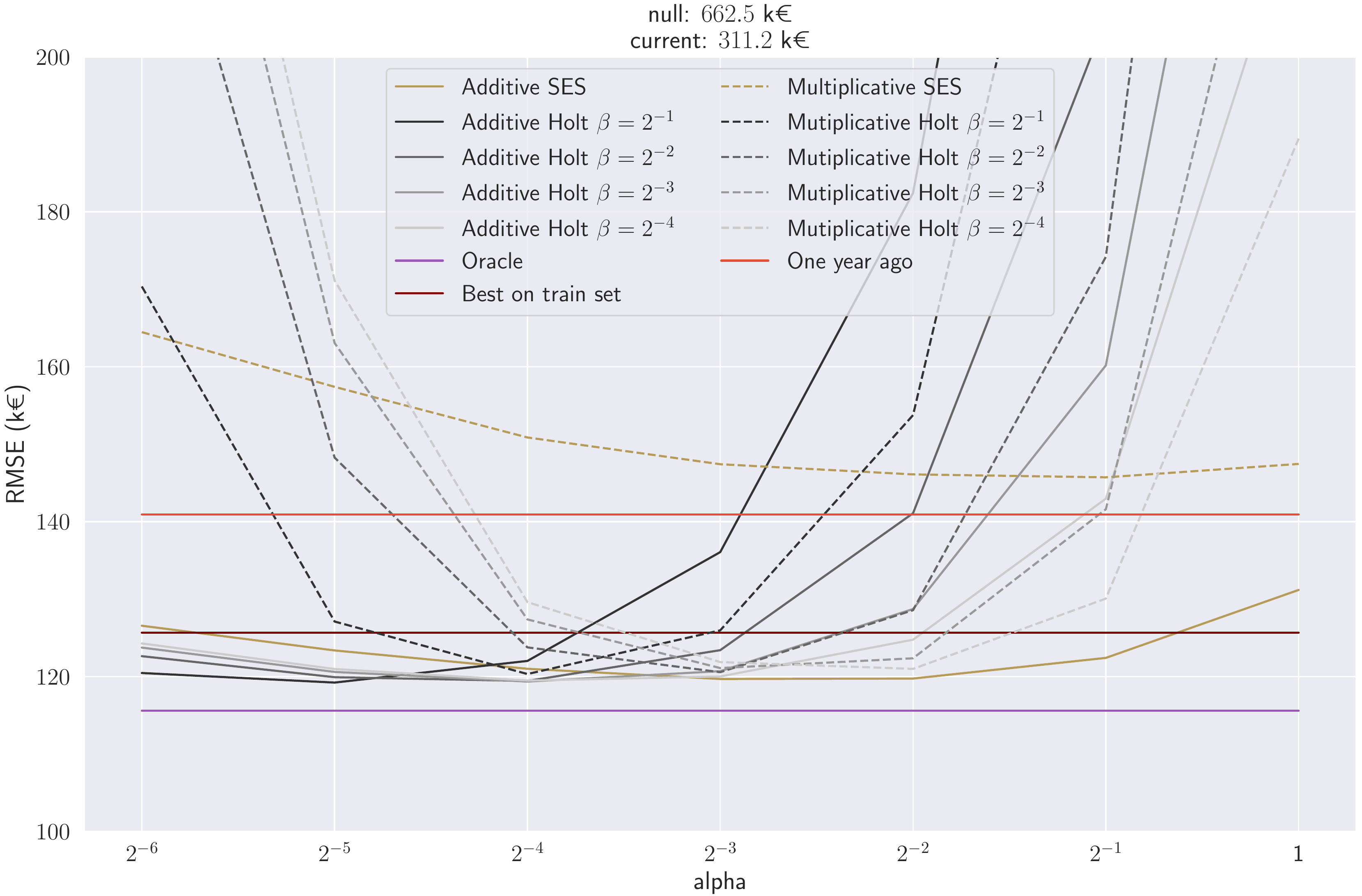}
\end{center}
\caption{\label{fig:elem} \small Performance [\emph{$y$--axis}, nominal scale] of the elementary forecasting methods
to forecast sales 6-week-ahead for 1 week (i.e., for $h=7$ and $n=1$),
in MAE [\emph{top figure}] and RMSE [\emph{bottom figure}],
depending on a tuning parameter $\alpha$ [\emph{$x$--axis}, logarithmic scale].
The acronym \texttt{SES} stands for simple exponential smoothing,
which can be implemented in a \texttt{Multiplicative} or an \texttt{Additive}
treatment of seasonality.
\texttt{Holt}'s linear trend method can also be implemented
with a \texttt{Multiplicative} or an \texttt{Additive} treatment of seasonality
and depends on a second parameter $\beta$: this is why
\texttt{Holt} elementary forecasting methods are instantiated
for several values of $\beta$. \smallskip \newline
The \texttt{Null}, \texttt{Current} and \texttt{One year ago}
elementary forecasting methods are the first three ones described in Section~\ref{sec:holt}
and do not depend on $\alpha$. The same can be said for the \texttt{Oracle} performance described in
Equation~\eqref{eq:guar-RMSE-joint} as well as
for the locally \texttt{Best on train set} predictor introduced in the beginning of Section~\ref{sec:perf-elem}.
The performance of \texttt{One year ago}, \texttt{Best on train set}, and \texttt{Oracle}
are therefore depicted by horizontal lines, while the one of
\texttt{Null} and \texttt{Current} can be found above the legend.
}
\end{figure}

We introduced three groups of elementary predictors in Section~\ref{sec:holt}.
The first group features the null predictor,
the predictor picking the sales achieved exactly one year ago,
and the predictor picking the current value of sales.
The second group features simple exponential smoothing
(which relies on a tuning parameter $\alpha \in [0,1]$),
with an additive or a multiplicative treatment of seasonality,
while the third group is made of Holt's linear trend method
(which relies on two tuning parameters $\alpha,\beta \in [0,1]$),
again with an additive or a multiplicative treatment of seasonality.
We pick a finite number of possible values for $\alpha$ and $\beta$
for our numerical experiments, namely:
\[
\alpha \in \bigl\{ 2^{-6}, \, 2^{-5}, \, 2^{-4}, \, 2^{-3}, \, 2^{-2}, \, 1/2, \, 1 \bigr\}
\qquad \mbox{and} \qquad
\beta \in \bigl\{ 2^{-4}, \, 2^{-3}, \, 2^{-2}, \, 1/2 \bigr\}
\]
(as the case $\beta = 1$ essentially corresponds to simple exponential smoothing).
As illustrated by Figure~\ref{fig:elem}, this leads to 73 elementary predictors:
$3$ predictors in the first group, $2 \times 7$ predictors based on simple exponential smoothing,
and $2 \times (7 \times 4)$ predictors based on Holt's linear trend method.

\paragraph{Definition of three meta-predictors.}
Based on these elementary predictors, we define
three meta-predictors: one legal meta-predictor and
two forward-looking ones (they ``cheat''
and use future data to pick among the elementary predictors).

The legal meta-predictor is to use at each node of the hierarchy
on the test set the elementary predictor that obtained the best performance
on the train set. We call this meta-predictor the locally best elementary
predictors on the train set; this is maybe the most natural meta-predictor in the eyes of practitioners.

A first forward-looking meta-predictor called the oracle prediction
was already defined in Section~\ref{sec:proj}:
it picks the locally best elementary predictors on the test set, that is,
with the notation of Section~\ref{sec:proj}, it achieves a performance in terms of
\begin{equation}
\label{eq:meta-pred1}
\frac{1}{|\Gamma|} \sum_{\gamma \in \Gamma} \min_{j=1,\ldots,J} \, \frac{1}{T} \sum_{t=1}^T \ell\Bigl(y_{t+h,\gamma},\wh{y}^{(j)}_{t+h,\gamma}\Bigr)\,,
\end{equation}
where $\ell$ is the loss function (absolute loss or squared loss) at hand.

Finally, we define a second forward-looking meta-predictor given by the globally best elementary predictor
on the test set, that is, the elementary predictor that obtains the best performance on the test set
when used on all nodes of the hierarchy; it achieves a performance in terms of
\begin{equation}
\label{eq:meta-pred2}
\min_{j=1,\ldots,J} \, \frac{1}{|\Gamma|} \sum_{\gamma \in \Gamma} \frac{1}{T} \sum_{t=1}^T \ell\Bigl(y_{t+h,\gamma},\wh{y}^{(j)}_{t+h,\gamma}\Bigr)\,,
\end{equation}

The notion of ``best'' depends on the underlying metric: MAE or RMSE. The globally best elementary predictor
on the test set may differ for each metric; the same can be said for locally best elementary predictors on the train or test set.

\paragraph{Figure~\ref{fig:elem}: Graphical comparison of these elementary predictors and meta-predictors.}
Figure~\ref{fig:elem} reports the performance of the elementary predictors and meta-predictors recalled or defined above,
in MAE and RMSE, for the case $(h,n) = (7,1)$, that is, for $6$--week-ahead forecasts relative to $1$ week of sales.
The four most interesting performance to read therein are, in order:
the predictor picking the sales achieved exactly one year ago (worst performance),
the locally best elementary predictors on the train set,
the globally best elementary predictor on the test set,
and the oracle (i.e., the locally best elementary predictors on the test set; best performance).
The best two such [meta-]predictors are forward-looking ones.
The gap between the locally best elementary predictors on the train set (the best legal meta-predictor) and
the globally best elementary predictor on the test set (a forward-looking meta-predictor)
is much larger in the case of RMSE than for MAE; it is almost null in the case of MAE.

As we show in the next sections, the performance of the aggregation algorithms considered
will be close to (but usually slightly larger than) the one of the globally best elementary predictor on the test set,
and in any case, significantly better than the one of the locally best elementary predictors on the train set.

\paragraph{Table~\ref{tab:meta-pred}: Numerical comparison of these elementary predictors and meta-predictors.}
The considerations above and the graphical comparison offered by
Figure~\ref{fig:elem} for the case $(h,n)=(7,1)$ show that our main indicators
are given by the performance of three meta-predictors:
the locally best elementary predictors on the train set (the legal meta-predictor),
the globally best elementary predictor on the test set (the first forward-looking meta-predictor),
and the oracle (i.e., the locally best elementary predictors on the test set; the second forward-looking
meta-predictor). Table~\ref{tab:meta-pred} reports these indicators in MAE and RMSE,
for various pairs $(h,n)$ of forecasting horizon $h-n$ and number $n$ of weeks to be forecast.

The main lessons are first that as expected, the farther away the horizon $h-n$,
the more important the average errors (in MAE or RMSE), and the larger the number $n$ of
weeks to be forecast, the smaller the average errors (a law-of-large-number probably smoothes out
sales when they are averaged over $n \geq 2$ weeks).
The number $n$ of weeks to forecast seems to have a greater impact on the average errors
than the horizon $h-n$, both for MAEs and RMSEs (actually, the RMSEs seem to be almost independent of the
horizon $h-n$).

Second, with one exception out of 18 cases of metric and $(h,n)$ pair considered,
the meta-predictors are consistently ranked, in terms of average errors,
as discussed above: the worst performance is achieved by the legal meta-predictor
(the locally best elementary predictors on the train set) and the best performance
is obtained by the oracle (the locally best elementary predictors on the test set),
with the other forward-looking meta-predictor (the globally best elementary predictor on the test set)
lying between them. The exception corresponds to the case of MAE and $(h,n)=(5,1)$.
This ranking may look surprising: the globally best elementary predictor on the test set
picks the same predictor at each node of the hierarchy and is less flexible than
the legal meta-predictor, that pick independently the elementary predictors at each node
(based on their performance on the train set). This probably means that the train set is much
different from the test set, which is probably due to highly non-stationary nature of e-commerce
data. This is why more flexible methods are welcome, like the online aggregation methods used in this article.

\begin{table}
\caption{\label{tab:meta-pred} Average errors (MAEs or RMSEs) in k€
for three meta-predictors (columns 5, 6, 7) depending on the metric considered
(column 1) as well as the horizon and number of weeks to be forecast (columns 2, 3, 4,
where we recall that the horizon is given by $h-n$) The three meta-predictors
considered were introduced in the beginning of Section~\ref{sec:perf-elem}:
the locally best elementary predictors on the train set (``Locally best on train set'', column~5),
the globally best elementary predictor on the test set (``Globally best on test set'', column~6),
and the oracle (which corresponds to the locally best elementary predictors on the test set,
abbreviated as ``Locally best on test set'', column~7).
.}
\begin{center}
\begin{tabular}{cccccccc}
\hline
Metric & Horizon & Group &  Pair $(h,n)$ & \phantom{esp} & Locally best & Globally best & Locally best \\
in k€ & & & & & on train set & on test set & on test set \\
& & & & & & & (= Oracle) \\
\hline
MAE & 6-week-ahead & for 1 week~ & $(7,1)$ & & $8.39$ & $8.30$ & $6.79$ \\
MAE & 6-week-ahead & for 2 weeks & $(8,2)$ & & $7.39$ & $7.34$ & $5.70$ \\
MAE & 6-week-ahead & for 4 weeks & $(10,4)$ & & $6.80$ & $6.59$ & $4.85$ \\
\hline
RMSE & 6-week-ahead & for 1 week~ & $(7,1)$ & & $125.68$ & $119.24$ & $115.59$ \\
RMSE & 6-week-ahead & for 2 weeks & $(8,2)$ & & $97.26$ & $90.94$ & $85.78$ \\
RMSE & 6-week-ahead & for 4 weeks & $(10,4)$ & & $82.36$ & $73.92$ & $68.23$ \\
\hline
MAE & 4-week-ahead & for 1 week~ & $(5,1)$ & & $8.18$ & $8.20$ & $6.78$ \\ 
MAE & 4-week-ahead & for 2 weeks & $(6,2)$ & & $7.27$ & $7.13$ & $5.66$ \\
MAE & 4-week-ahead & for 4 weeks & $(8,4)$ & & $6.58$ & $6.34$ & $4.77$ \\
\hline
RMSE & 4-week-ahead & for 1 week~ & $(5,1)$ & & $126.04$ & $119.94$ & $117.12$ \\
RMSE & 4-week-ahead & for 2 weeks & $(6,2)$ & & $98.31$ & $90.79$ & $85.84$ \\
RMSE & 4-week-ahead & for 4 weeks & $(8,4)$ & & $79.69$ & $72.82$ & $67.67$ \\
\hline
MAE & 1-week-ahead & for 1 week~ & $(2,1)$ & & $7.63$ & $7.42$ & $6.42$ \\
MAE & 1-week-ahead & for 2 weeks & $(3,2)$ & & $6.69$ & $6.48$ & $5.42$ \\
MAE & 1-week-ahead & for 4 weeks & $(5,4)$ & & $6.18$ & $5.90$ & $4.60$ \\
\hline
RMSE & 1-week-ahead & for 1 week~ & $(2,1)$ & & $124.36$ & $118.17$ & $115.23$ \\
RMSE & 1-week-ahead & for 2 weeks & $(3,2)$ & & $95.48$ & $89.37$ & $85.11$ \\
RMSE & 1-week-ahead & for 4 weeks & $(5,4)$ & & $78.86$ & $71.35$ & $66.90$ \\
\hline
\end{tabular}
\end{center}
\end{table}

\paragraph{Discussion on the two metrics considered: MAE and RMSE.}
We add a final note on the orders of magnitude between MAEs and RMSEs. They differ by a factor of 10 to 15,
with the RMSEs being roughly 10 to 15 times larger than the MAEs.
This is because RMSEs are extremely sensitive to extreme values. These extreme values may
correspond, in e-commerce, to external interferences (sales periods, disruptions in supply of some products,
launches of new products, crises: financial, sanitary, social crises). The impact of such external interferences
needs to be forecast separately, with ad hoc models and methods. The scope of the present article
is therefore rather on forecasting sales in stationary regimes, that is, for ``ordinary'' or routine circumstances.
And in such regimes and circumstances, extreme values are rare and less important in the eyes of the decision-makers
than typical values. This is why, in the sequel, while reporting both MAEs and RMSEs, we will be more
interested in the performance in MAE.

\clearpage

\subsection{Average Performance of the Aggregation Algorithms: MAE, RMSE}
\label{sec:res:main}

Now that we identified some benchmark performance,
we may compare the performance of the aggregation algorithms considered
to this performance. We proceed in two steps: first, we tabulate the performance
of these algorithms under their various specifications on a given case, namely, $(h,n)=(7,1)$
corresponding to 6-week-ahead forecasting of 1 week of sales. We show that
they all achieve a rather similar performance. For the second part of the study,
we thus set (somewhat arbitrarily) a given algorithm with given specifications, namely, ML-Poly with absolute loss,
without the gradient trick and with projection, and tabulate its performance
depending on $(h,n)$, that is, depending on the forecasting horizon $h-n$ and
the number $n$ of weeks to be forecast.

\paragraph{First part: Little impact of the algorithm picked and of its specifications.}
As explained above and as is summarized in Table~\ref{tab:perf-algo}, we consider
three algorithms under $2 \times 2 \times 3 = 12$ possible specifications
(given by choices made for the loss function, gradient trick, and projection step).
We report the performance of each specification in MAE and RMSE for the case $(h,n) = (7,1)$.

In terms of RMSEs, the various algorithms and specifications thereof (with one exception) are
virtually undistinguishable, with RMSEs all around 120 k€ when the gradient
trick is not applied (and slightly larger, up to 125.4 k€ when it is applied).
The Euclidean projection barely improves the RMSE (Section~\ref{sec:proj} recalls
why this projection must improve the RMSE).
The exception to the virtually undistinguishable performance is ML-Prod without the gradient trick,
which fares much worse than ML-Prod with the gradient trick
or the various specifications of ML-Poly and BOA.

A summary of the same kind may be written for MAEs: many of the algorithms and specifications
thereof have MAEs around 8 k€ (slightly larger values are suffered when the Euclidean
projection is applied). The projection in absolute norm slightly worsens the results
(Section~\ref{sec:proj} recalls why this projection came with no positive guarantee on
its impact on the MAE). The loss function $\ell$ and the gradient trick
have little impact, though the absolute loss seems a slightly better choice than the
square loss, and though it seems better not to resort to the gradient trick.

The conclusion from this study is that the choice of the specific aggregation algorithm
and of its specification is not of utmost importance. For the rest of the study,
we will fix an algorithm (namely, ML-Poly) with the simplest specification:
no gradient trick, no projection, and absolute loss (which is in line with our focus
on MAE). The BOA algorithm under this simplest specification gets a better performance
on the case $(h,n)=(7,1)$ but we have a personal preference for ML-Poly, which
was designed by one of the co-authors of this article.

\paragraph{Second part: Relative performance compared to the meta-predictors.}
Table~\ref{tab:perf-algo-hn} studies the performance of a given algorithm
under a given specification, namely, ML-Poly with absolute loss, no gradient trick, no projection,
as concluded from the paragraphs above. It compares its performance
to the one of the three meta-predictors discussed in Section~\ref{sec:perf-elem}.
Two main benchmarks were outlined in the latter section:
the locally best predictors on the train set (which is a legal meta-predictor)
and the globally best predictor on the test set (which is a forward-looking meta-predictor).

%

The aggregation algorithm consistently outperforms the locally best predictors on the train set,
for all cases $(h,n)$ of horizon $h-n$ and number $n$ of weeks to be forecast, both in MAE and RMSE.
The improvement is typically around $5\%$ (it ranges from a minimal $3.6\%$ to a maximal $7.1\%$
relative improvement). We recall that the the locally best predictors on the train set
actually depend on the underlying metric: MAE or RMSE.

The situation is two-fold for the comparison of the aggregation algorithm to
the globally best predictor on the test set: the former consistently
outperforms the latter in our favorite metric, namely, MAE, with relative improvements
in the range $2.0\%$--$4.5\%$.
On the opposite, the aggregation algorithm is consistently
outperformed by the globally best predictor on the test set in RMSE,
within a $0.5\%$--$5.3\%$ range.

Table~\ref{tab:perf-algo-hn2} studies the performance of the
same algorithm, ML-Poly, under a slightly different specification: still without a gradient trick and without projection,
but with square loss instead of absolute loss. This should favor RMSE performance.
The picture is about the same: consistent improvement in performance
over the locally best predictors on the train set (with range $2.3\%$--$8.3$\%);
mixed pictures for the comparison to the globally best predictor on the test set,
and indeed, the RMSE performance is globally improved.

However, given that our aim is to predict ``ordinary'' (and not extreme) values,
we are more interested in the MAE performance. For MAE performance, the aggregation algorithm
considered in Table~\ref{tab:perf-algo-hn}
is consistently better than the forward-looking meta-predictor picking the globally best predictor on the test set
(on all 9 cases). For the one of Table~\ref{tab:perf-algo-hn2} (for which we changed the loss function into
square loss), the improvement holds for 7 out of 9 cases (and for the 2 other ones, the difference in
performance is negligible, smaller than~$0.4$\%).

\vfill

\begin{table}[h]
\caption{\label{tab:perf-algo} Average errors in k€ (in MAE, columns 4--6, and in
RMSE, columns 7--9) for the case $(h,n) = (7,1)$, that is, for 6-week-ahead forecasts of 1 week of sales,
for the three algorithms considered (ML-Poly, rows 1--4; BOA, rows 5--8;
ML-Prod, rows 9--12), under various specifications thereof:
loss function used (see Section~\ref{sec:aggreg2}),
either the absolute loss $|\,\cdot\,|$ or the square loss $(\,\,\cdot\,\,)^2$,
as indicated in column~3;
whether the gradient trick (see Section~\ref{sec:aggreg3}) is applied
or not (column~2, ``yes'' or ''no''); whether a projection step (see Section~\ref{sec:proj}) is added
or not (``no proj.'', columns~4 and~7), and if so, whether a projection in Euclidean norm (``L2--proj.'', columns~5 and~8) or in
absolute norm (``L1--proj.'', columns~6 and~9) is used. }
\begin{center}
\begin{tabular}{lclcccccccc}
\hline
\multicolumn{11}{c}{Case $(h,n)=(7,1)$, i.e., 6-week-ahead forecasts for 1 week} \\
\hline
Algor. & Gradient & Loss & \phantom{e} & \multicolumn{3}{c}{MAE in k€} & \phantom{ep} & \multicolumn{3}{c}{RMSE in k€} \\
& trick & ~~$\ell$ & & no proj. & L2--proj. & L1--proj. & & no proj. & L2--proj. & L1--proj. \\
\hline
ML-Poly & no & $|\,\cdot\,|$ & & $7.97$ & $8.80$ & $8.07$ & & $120.39$ & $120.31$ & $120.03$\\
ML-Poly & no & $(\,\,\cdot\,\,)^2$ & & $8.04$ & $8.85$ & $8.10$ & & $119.79$ & $119.71$ & $119.39$\\
ML-Poly & yes & $|\,\cdot\,|$ & & $8.05$ & $8.79$ & $8.11$ & & $121.90$ & $121.79$ & $121.25$\\
ML-Poly & yes & $(\,\,\cdot\,\,)^2$ & & $8.26$ & $8.91$ & $8.35$ & & $125.40$ & $125.31$ & $124.90$\\
\hline
ML-Prod & no & $|\,\cdot\,|$ & & $13.28$ & $16.26$ & $13.06$ & & $278.45$ & $278.04$ & $226.17$ \\
ML-Prod & no & $(\,\,\cdot\,\,)^2$ & & $12.99$ & $16.39$ & $12.62$ & & $277.51$ & $276.98$ & $215.62$ \\
ML-Prod & yes & $|\,\cdot\,|$ & & $7.94$ & $8.49$ & $8.01$ & & $120.54$ & $120.48$ & $120.32$ \\
ML-Prod & yes & $(\,\,\cdot\,\,)^2$ & & $8.10$ & $8.64$ & $8.14$ & & $120.17$ & $120.12$ & $119.98$ \\
\hline
BOA & no & $|\,\cdot\,|$ & & $7.93$ & $8.72$ & $8.07$ & & $120.39$ & $120.30$ & $120.94$\\
BOA & no & $(\,\,\cdot\,\,)^2$ & & $8.06$ & $8.68$ & $8.17$ & & $120.44$ & $120.37$ & $120.89$\\
BOA & yes & $|\,\cdot\,|$ & & $7.97$ & $8.79$ & $8.04$ & & $121.73$ & $121.61$ & $121.45$\\
BOA & yes & $(\,\,\cdot\,\,)^2$ & & $8.12$ & $8.70$ & $8.21$ & & $122.60$ & $122.53$ & $122.64$\\
\hline
\end{tabular}
\end{center}
\end{table}

\clearpage

\begin{table}
\caption{\label{tab:perf-algo-hn} Average errors in k€ (columns 3--6)
and relative differences in errors (columns 7--8) for a given aggregation algorithm under a given specification
(namely, ML-Poly with the absolute loss $\ell$, no gradient trick, no projection),
depending on the pairs $(h,n)$ and on the metrics (MAE or RMSE) considered (see columns 1--2).
The same cases and meta-predictors as in Table~\ref{tab:meta-pred} are tabulated:
columns~1--4 and~6 are exactly equal to the corresponding columns in Table~\ref{tab:meta-pred}.
Column~5 reports the absolute performance of the aggregation algorithm considered (``Aggregation'', with short-hand ``Aggreg'').
Columns~7 and~8 report the relative performance of the aggregation algorithm considered,
compared either to the locally best predictors on the train set (short-hand ``Loc-Train'', column~7)
or to the globally best predictor on the test set (short-hand ``Glob-Test'', column~8).
Negative (respectively, positive) numbers in columns~7 and~8 indicate that the performance of the aggregation algorithm
is better (respectively, worse) than to the meta-predictor it is compared with. The line titled ``Legal meta-predictor''
recalls which meta-predictors are legal (i.e., only rely on information available at the time they
issues their forecasts, ``Yes'') and which of them are using future data (``No'').}
\bigskip

\hspace{-.5cm}
{\small
\begin{tabular}{cccccccccc}
\hline
\multicolumn{10}{c}{Algorithm ML-Poly, with specifications: $\ell$ is the absolute loss, no gradient trick, no projection} \\
\hline
Metric & Pair & \phantom{es} & Locally best & Globally best & Aggregation & Locally best & \phantom{es} & Aggreg. & Aggreg.\ \\
in k€ & $(h,n)$ & & on train set & on test set & & on test set & & vs. & vs. \\
& & & (= Loc-Train) & (= Glob-Test) & (= Aggreg) & (= Oracle) & & Loc-Train & Glob-Test \\
\hline
\multicolumn{3}{l}{Legal meta-predictor} & Yes & No & Yes & No \\
\hline
 MAE & $(7,1)$ & & $8.39$ & $8.30$ & $7.97$  & $6.79$ & & $-5.1$\% & $-4.0\%$ \\
 MAE & $(8,2)$ & & $7.39$ & $7.34$ & $7.12$  & $5.70$ & & $-3.6$\% & $-3.1\%$ \\
 MAE & $(10,4)$ & & $6.80$ & $6.59$ & $6.42$  & $4.85$ & & $-5.5$\% & $-2.5\%$\\
\hline
 RMSE & $(7,1)$ & & $125.68$  & $119.24$ &  $120.39$  & $115.59$ & & $-4.2$\% & $+1.0\%$\\
 RMSE & $(8,2)$ & & $97.26$  & $90.94$   &  $93.62$  & $85.78$ & & $-3.7$\% & $+3.0\%$ \\
 RMSE & $(10,4)$ & & $82.36$ & $73.92$  & $78.02$  & $68.23$ & &$-5.3$\% & $+5.3\%$ \\
\hline
 MAE & $(5,1)$ & & $8.18$ & $8.20$     & $7.83$  & $6.78$ & &$-4.4$\% & $-4.5\%$ \\
 MAE & $(6,2)$ & & $7.27$ & $7.13$    & $6.83$  & $5.66$ & & $-6.0$\% & $-4.2\%$ \\
 MAE & $(8,4)$ & & $6.58$ & $6.34$   & $6.21$  & $4.77$ & & $-5.6$\% & $-2.0\%$ \\
\hline
 RMSE & $(5,1)$ & & $126.04$ & $119.94$   & $120.84$  & $117.12$ & & $-4.1$\% & $+0.8\%$ \\
 RMSE & $(6,2)$ & & $98.31$ & $90.79$   & $92.46$  & $85.84$ & & $-6.0$\% & $+1.8\%$ \\
 RMSE & $(8,4)$ & & $79.69$ & $72.82$   & $75.60$  & $67.67$ & & $-5.1$\% & $+3.8\%$ \\
\hline
 MAE & $(2,1)$ & & $7.63$ & $7.42$     & $7.11$  & $6.42$ & & $-6.8$\% & $-4.2\%$ \\
 MAE & $(3,2)$ & & $6.69$ & $6.48$   & $6.30$  & $5.42$ & & $-5.8$\% & $-2.8\%$ \\
 MAE & $(5,4)$ & & $6.18$ & $5.90$   & $5.74$  & $4.60$ & & $-7.1$\% & $-2.7\%$ \\
\hline
 RMSE & $(2,1)$ & & $124.36$  & $118.17$  & $118.71$ & $115.23$  & & $-4.6$\%  & $+0.5$\%  \\
 RMSE & $(3,2)$ & & $95.48$  & $89.37$  & $90.17$  & $85.11$  & & $-5.6$\%  & $+0.9$\%  \\
 RMSE & $(5,4)$ & & $78.86$  & $71.35$  & $73.82$  & $66.90$  & & $-6.4$\%  & $+3.5$\%  \\
\hline
\end{tabular}
}
\end{table}

\begin{table}
\caption{\label{tab:perf-algo-hn2}
Same content as in Table~\ref{tab:perf-algo-hn}, still with ML-Poly as an aggregation
algorithm, but under a slightly different specification:
no gradient trick, no projection (as in Table~\ref{tab:perf-algo-hn}), but with the square loss $\ell$
(instead of the absolute loss as in Table~\ref{tab:perf-algo-hn}).
Only the values of columns~5, 7, 8 differ from the ones of Table~\ref{tab:perf-algo-hn}.}
\bigskip

\hspace{-.5cm}
{\small
\begin{tabular}{cccccccccc}
\hline
\multicolumn{10}{c}{Algorithm ML-Poly, with specifications: $\ell$ is the square loss, no gradient trick, no projection} \\
\hline
Metric & Pair & \phantom{es} & Locally best & Globally best & Aggregation & Locally best & \phantom{es} & Aggreg. & Aggreg.\ \\
in k€ & $(h,n)$ & & on train set & on test set & & on test set & & vs. & vs. \\
& & & (= Loc-Train) & (= Glob-Test) & (= Aggreg) & (= Oracle) & & Loc-Train & Glob-Test \\
\hline
\multicolumn{3}{l}{Legal meta-predictor} & Yes & No & Yes & No \\
\hline
MAE & $(7,1)$ & & $8.39$  & $8.30$  & $8.04$  & $6.79$  & & $-4.2$\%  & $-3.2$\%  \\
MAE & $(8,2)$ & & $7.39$  & $7.34$  & $7.22$  & $5.70$  & & $-2.3$\%  & $-1.7$\%  \\
MAE & $(10,4)$ & & $6.80$  & $6.59$  & $6.61$  & $4.85$  & & $-2.8$\%  & $+0.4$\%  \\
\hline
RMSE & $(7,1)$ & & $125.68$  & $119.24$  & $119.79$  & $115.59$  & & $-4.7$\%  & $+0.5$\%  \\
RMSE & $(8,2)$ & & $97.26$  & $90.94$  & $93.23$  & $85.78$  & & $-4.1$\%  & $+2.5$\%  \\
RMSE & $(10,4)$ & & $82.36$  & $73.92$  & $77.44$  & $68.23$  & & $-6.0$\%  & $+4.8$\%  \\
\hline
MAE & $(5,1)$ & & $8.18$  & $8.20$  & $7.90$  & $6.78$  & & $-3.5$\%  & $-3.7$\%  \\
MAE & $(6,2)$ & & $7.27$  & $7.13$  & $6.86$  & $5.66$  & & $-5.7$\%  & $-3.9$\%  \\
MAE & $(8,4)$ & & $6.58$  & $6.34$  & $6.36$  & $4.77$  & & $-3.3$\%  & $+0.3$\%  \\
\hline
RMSE & $(5,1)$ & & $126.04$  & $119.94$  & $121.55$  & $117.12$  & & $-3.6$\%  & $+1.3$\%  \\
RMSE & $(6,2)$ & & $98.31$  & $90.79$  & $91.53$  & $85.84$  & & $-6.9$\%  & $+0.8$\%  \\
RMSE & $(8,4)$ & & $79.69$  & $72.82$  & $75.62$  & $67.67$  & & $-5.1$\%  & $+3.8$\%  \\
\hline
MAE & $(2,1)$ & & $7.63$  & $7.42$  & $7.18$  & $6.42$  & & $-5.9$\%  & $-3.2$\%  \\
MAE & $(3,2)$ & & $6.69$  & $6.48$  & $6.28$  & $5.42$  & & $-6.2$\%  & $-3.2$\%  \\
MAE & $(5,4)$ & & $6.18$  & $5.90$  & $5.79$  & $4.60$  & & $-6.2$\%  & $-1.8$\%  \\
\hline
RMSE & $(2,1)$ & & $124.36$  & $118.17$  & $118.93$  & $115.23$  & & $-4.4$\%  & $+0.6$\%  \\
RMSE & $(3,2)$ & & $95.48$  & $89.37$  & $89.17$  & $85.11$  & & $-6.6$\%  & $-0.2$\%  \\
RMSE & $(5,4)$ & & $78.86$  & $71.35$  & $72.33$  & $66.90$  & & $-8.3$\%  & $+1.4$\%  \\
\hline
\end{tabular}
}
\end{table}

\clearpage
\subsection{An Intrinsic Evaluation of Performance: Mean Percentages of Error}
\label{sec:mape}

So far, we have been discussing performance in MAE or RMSE and needed
benchmarks to assess the quality of the forecasts issued by the aggregation algorithms
(and the latter outperformed these benchmarks: the locally best predictors on the train set and
the globally best predictor on the test set). Put differently, we were only discussing relative performance. We now want to move
to a more intrinsic evaluation of the performance of the aggregation algorithms (and of the
meta-predictors). To that end, we use a mean absolute percentage of error [MAPE] as our criterion.
The latter is not so easy to define, as many sales $y_{t,\gamma}$ are null (see Section~\ref{sec:descrdata}), and therefore,
the classical definition
\[
\cancel{\frac{1}{T} \sum_{t=1}^T \sum_{\gamma \in \Gamma} \frac{\bigl| y_{t+h,\gamma} - \wh{y}_{t+h,\gamma} \bigr|}{y_{t+h,\gamma}}}
\]
fails. This is why we adapt this classical definition of MAPE to our needs, as follows.
We provide this adaptation for a given a subset $\Gamma_{\sub} \subseteq \Gamma$ of nodes
(sometimes $\Gamma_{\sub}$ will be the set $\Gamma$ of all nodes, and sometimes a strict subset, e.g.,
given by all subsubfamilies):
\[
\mape = \frac{1}{T} \sum_{t=1}^T \frac{\displaystyle{\sum_{\gamma \in \Gamma_{\sub}} \bigl| y_{t+h,\gamma} - \wh{y}_{t+h,\gamma} \bigr|}}{\displaystyle{\sum_{\gamma \in \Gamma_{\sub}} y_{t+h,\gamma}}}
\]
When the subset $\Gamma_{\sub}$ is large enough (whenever it contains a significant number of nodes),
the denominator is positive and the MAPE is well defined in this way.

\subsubsection{On the Entire Hierarchy $\Gamma$}

\begin{figure}[p]
\begin{center}
\includegraphics[width=\textwidth]{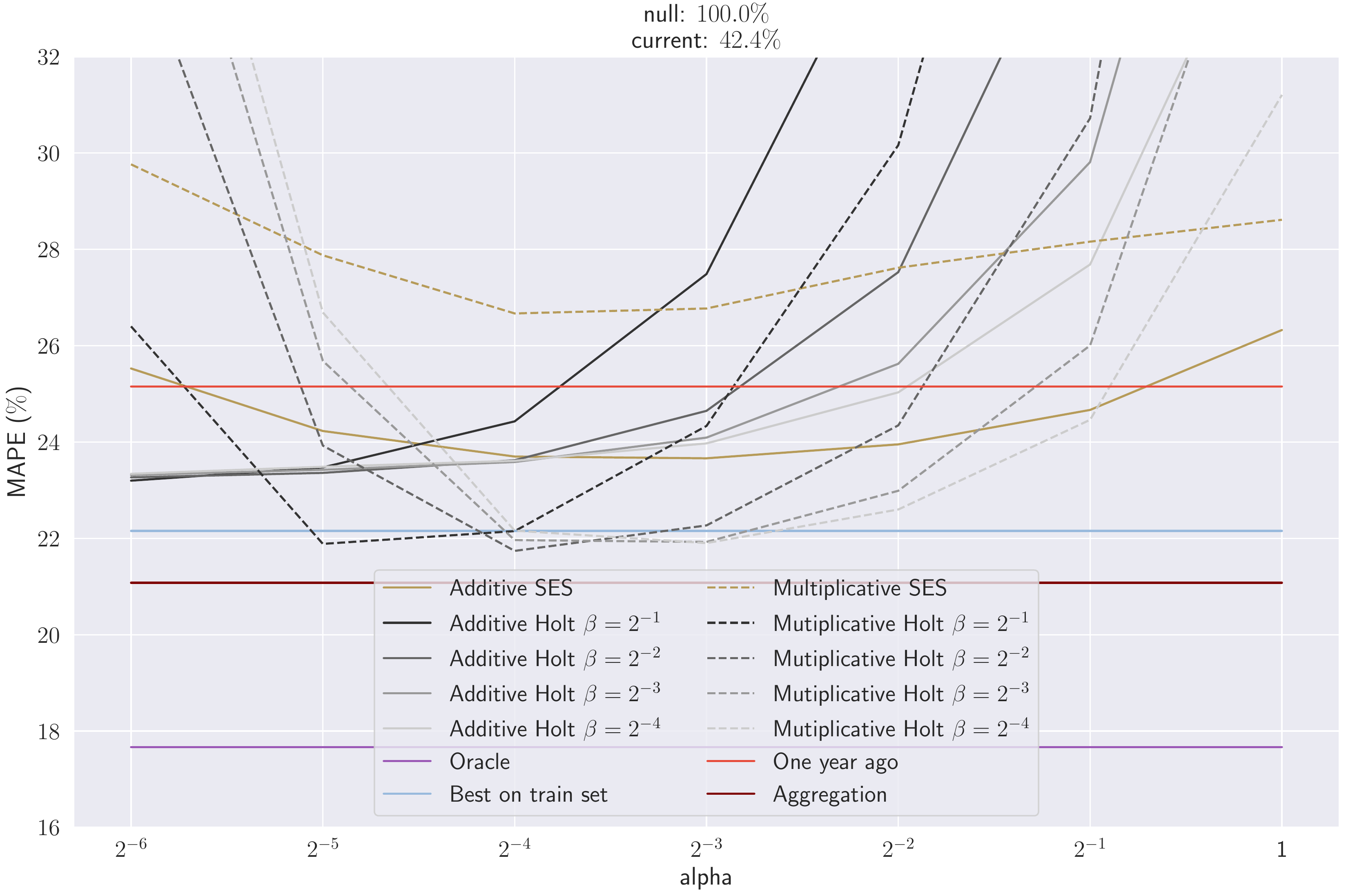} \vspace{-1cm} \\
\end{center}
\caption{\label{fig:mape} Performance in MAPE [\emph{$y$--axis}, in $\%$] of the elementary forecasting methods,
of some meta-predictors, and of a given aggregation algorithm
to forecast sales 6-week-ahead for 1 week (i.e., for $h=7$ and $n=1$),
depending on a tuning parameter $\alpha$ [\emph{$x$--axis}, logarithmic scale].
The same acronyms are used as in Figure~\ref{fig:elem}, with the addition
of an \texttt{Aggregation} algorithm, namely,
ML-Poly with the absolute value, no gradient trick and no projection;
its performance is independent of $\alpha$ and is therefore depicted by an horizontal
line.}
\end{figure}

We first discuss global performance, on the entire hierarchy of nodes $\Gamma$.
Figure~\ref{fig:mape} and Table~\ref{tab:perf-algo-mape} are counterparts of similar figures and
a similar table in the case of MAE and RMSE.
They display graphically (Figure~\ref{fig:mape}) the performance in MAPE
of the elementary predictors and meta-predictors introduced in Section~\ref{sec:perf-elem},
as well as the one of a given aggregation algorithm, namely,
ML-Poly with the absolute loss, no gradient trick, no projection
(just as in Section~\ref{sec:res:main} above).
Of course, all meta-predictors defined in terms of a ``best predictor''
or ``best predictors'' as in \eqref{eq:meta-pred1} or~\eqref{eq:meta-pred2}are defined with respect
to the loss function
\begin{equation}
\label{eq:bestGamma}
\ell\Bigl(y_{t+h,\gamma},\wh{y}^{(j)}_{t+h,\gamma}\Bigr)
= \frac{\Bigl| y_{t+h,\gamma} - \wh{y}^{(j)}_{t+h,\gamma} \Bigr|}{\displaystyle{\sum_{g \in \Gamma} y_{t+h,g}}}\,.
\end{equation}
(We should actually add arguments to $\ell$, as the loss computed depends on all observations $y_{t+h,g}$, not
just the one at the node $\gamma$.)

In terms of relative performance, Figure~\ref{fig:mape} and Table~\ref{tab:perf-algo-mape}
for MAPE show a similar ranking as Figure~\ref{fig:elem} and Table~\ref{tab:perf-algo-hn} for MAE:
the aggregation algorithm consistently outperforms the locally best predictors on the train set
and the globally best predictor on the test set.

We are more interested in an intrinsic evaluation of the performance,
which is why we considered MAPE in the first place. The MAPEs of the aggregation
algorithm lie between $15.24\%$ and $21.08\%$ (these MAPEs are larger when the horizon is farther away
and/or the number of weeks to be forecast is smaller). This is a nice performance,
but we break it down by levels of the hierarchy before issuing any deeper comments.

\begin{table}[p]
\caption{\label{tab:perf-algo-mape} Performance in MAPE (columns 3--6)
and relative differences in MAPE (columns 7--8) for a given aggregation algorithm under a given specification
(namely, ML-Poly with the absolute loss $\ell$, no gradient trick, no projection),
depending on the pairs $(h,n)$. The same structure and conventions are used as for Table~\ref{tab:perf-algo-hn}.}
\bigskip

\hspace{-0.5cm}
{\small
\begin{tabular}{cccccccccc}
\hline
\multicolumn{10}{c}{Algorithm ML-Poly, with specifications: $\ell$ is the absolute loss, no gradient trick, no projection} \\
\hline
Metric & Pair & \phantom{es} & Locally best & Globally best & Aggregation & Locally best & \phantom{es} & Aggreg. & Aggreg.\ \\
& $(h,n)$ & & on train set & on test set & & on test set & & vs. & vs. \\
& & & (= Loc-Train) & (= Glob-Test) & (= Aggreg) & (= Oracle) & & Loc-Train & Glob-Test \\
\hline
\multicolumn{3}{l}{Legal meta-predictor} & Yes & No & Yes & No \\
\hline
 MAPE & $(7,1)$ & & $22.15$  & $21.74$  & $21.08$  & $17.67$  & & $-4.9$\%  & $-3.0$\%  \\
 MAPE & $(8,2)$ & & $19.46$  & $19.04$  & $18.74$  & $14.69$  & & $-3.7$\%  & $-1.6$\%  \\
 MAPE & $(10,4)$ & & $18.21$  & $17.46$  & $17.19$  & $12.86$  & & $-5.6$\%  & $-1.6$\%  \\
\hline
 MAPE & $(5,1)$ & & $21.43$  & $21.34$  & $20.50$  & $17.56$  & & $-4.3$\%  & $-3.9$\%  \\
 MAPE & $(6,2)$ & & $18.94$  & $18.47$  & $17.89$  & $14.53$  & & $-5.6$\%  & $-3.1$\%  \\
 MAPE & $(8,4)$ & & $17.57$  & $16.75$  & $16.66$  & $12.59$  & & $-5.2$\%  & $-0.5$\%  \\
\hline
 MAPE & $(2,1)$ & & $19.76$  & $19.33$  & $18.45$  & $16.63$  & & $-6.7$\%  & $-4.6$\%  \\
 MAPE & $(3,2)$ & & $17.53$  & $16.76$  & $16.32$  & $13.90$  & & $-6.9$\%  & $-2.7$\%  \\
 MAPE & $(5,4)$ & & $16.53$  & $15.57$  & $15.24$  & $12.09$  & & $-7.8$\%  & $-2.1$\%  \\
\hline
\end{tabular}
}
\end{table}

\subsubsection{Level by Level}

We now explore MAPE performance by levels of the hierarchy:
by taking subsets $\Gamma_{\sub}$ given by all subsubfamilies,
or by all subfamilies, or by all families. We also report the MAPE
for predicting the total sales, i.e., $\Gamma_{\sub}$ is the root-node singleton:
$\Gamma_{\sub} = \{\mbox{root}\}$.
When considering a ``best predictor'' or ``best predictors'' for our meta-predictors,
similarly to the definition given by~\eqref{eq:bestGamma}, by updating the summation in the denominator of the latter,
we resort to the loss function
\begin{equation}
\ell\Bigl(y_{t+h,\gamma},\wh{y}^{(j)}_{t+h,\gamma}\Bigr)
= \frac{\Bigl| y_{t+h,\gamma} - \wh{y}^{(j)}_{t+h,\gamma} \Bigr|}{\displaystyle{\sum_{g \in \Gamma_{\sub}} y_{t+h,g}}}\,.
\end{equation}
Put differently, ``best'' is now in terms of MAPE and of the considered level of the hierarchy.

Results are reported in Table~\ref{tab:perf-algo-mape-level}.
We first discuss the intrinsic performance of the aggregation algorithm:
it obtains an MAPE of about $32\%$ on the case of all subsubfamilies, which is the most important case to consider.
Indeed, the volumes of sales at this level are then broken down into specific products, either existing ones
or new products to be launched. The forecasts at this level support and drive the decision-making.
This $32\%$ MAPE is comparable to MAPEs observed for the forecasting of sales in retail distribution.

The MAPE performance of course improves as we go up in the hierarchy: it equals about $22\%$ for
all subfamilies, $18\%$ for all families, and $12\%$ for the root note. We recall that the MAPE
performance for the entire hierarchy (i.e., putting together all levels) equals about $21\%$.

Now, in terms of relative performance (i.e., when the aggregation algorithm is compared
to meta-predictors), we observe that the aggregation algorithm consistently outperforms
the legal meta-predictor given by the locally best predictors on the train set,
while it outperforms the forward-looking meta-predictor given the globally best predictor on the test
set on the two cases that are of most interest for us: all subsubfamilies, and the entire hierarchy;
it is outperformed by that forward-looking meta-predictor on the three other cases:
root node (also known as total node), all families, all subfamilies (very slightly).

\vfill
\begin{table}[h]
\caption{\label{tab:perf-algo-mape-level} Performance in MAPE (columns 2--5)
and relative differences in MAPE (columns 6--7) for a given aggregation algorithm under a given specification
(namely, ML-Poly with the absolute loss $\ell$, no gradient trick, no projection),
depending on the hierarchy level(s) considered. The line ``Entire hierarchy'' corresponds
to taking $\Gamma_{\sub} = \Gamma$, while the four other lines correspond
each to an element of a partition of $\Gamma$ by levels:
$\Gamma_{\sub} = \{\mbox{root}\}$ for the line ``Total node'',
$\Gamma_{\sub}$ the subsets of all families, subfamilies, subsubfamilies, respectively.
A similar structure of the results as for Table~\ref{tab:perf-algo-hn} is used.}
\bigskip

\hspace{-0.5cm}
{\small
\begin{tabular}{lccccccccc}
\hline
\multicolumn{9}{c}{MAPE; case $(h,n)=(7,1)$; algorithm ML-Poly, with specifications:} \\
\multicolumn{9}{c}{$\ell$ is the absolute loss, no gradient trick, no projection} \\
\hline
& \phantom{e} & Locally best & Globally best & Aggregation & Locally best & \phantom{e} & Aggreg. & Aggreg.\ \\
Level & & on train set & on test set & & on test set & & vs. & vs. \\
& & (= Loc-Train) & (= Glob-Test) & (= Aggreg) & (= Oracle) & & Loc-Train & Glob-Test \\
\hline
\multicolumn{2}{l}{Legal} & Yes & No & Yes & No \\
\hline
Entire hierarchy & & $22.15\%$ & $21.74\%$ & $21.08\%$ & $17.67\%$ & & $-4.9\%$ & $-3.0\%$ \\
Total node & & $12.46\%$ & $11.34\%$ &  $11.71\%$ & $11.34\%$ & & $-6.0\%$ & $+3.3\%$ \\
Families & & $18.70\%$ & $16.96\%$ & $18.32\%$ & $15.56\%$ & & $-2.0\%$ & $+8.0\%$ \\
Subfamilies & & $23.46\%$ & $22.11\%$ & $22.20\%$ & $18.49\%$ & & $-5.4\%$ & $+0.4\%$ \\
Subsubfamilies & & $33.99\%$ & $36.00\%$ & $32.09\%$ & $25.29\%$ & & $-5.6\%$ & $-10.9\%$ \\
\hline
\end{tabular}
}
\end{table}

%
%
%
%

\clearpage
\subsection{Beyond Average Performance}
\label{sec:robustness}

We go beyond average performance measures in this section
and illustrate that the performance of the aggregation algorithms
is not only better on average but everywhere, compared to, e.g., the natural benchmark given by
the locally best predictors on the train set.
To do so, we consider the absolute errors suffered for predicting the sales of each of the
3,004 subsubfamilies on each of the 52 weeks of the test set, which leads to
$52 \times 3,\!004 = 156,\!208$ absolute errors. We do so for the case $(h,n)=(7,1)$,
i.e., for 6-week-ahead-forecasting of 1 week of sales.

\begin{figure}[h]
\begin{center}
\includegraphics[width=0.88\textwidth]{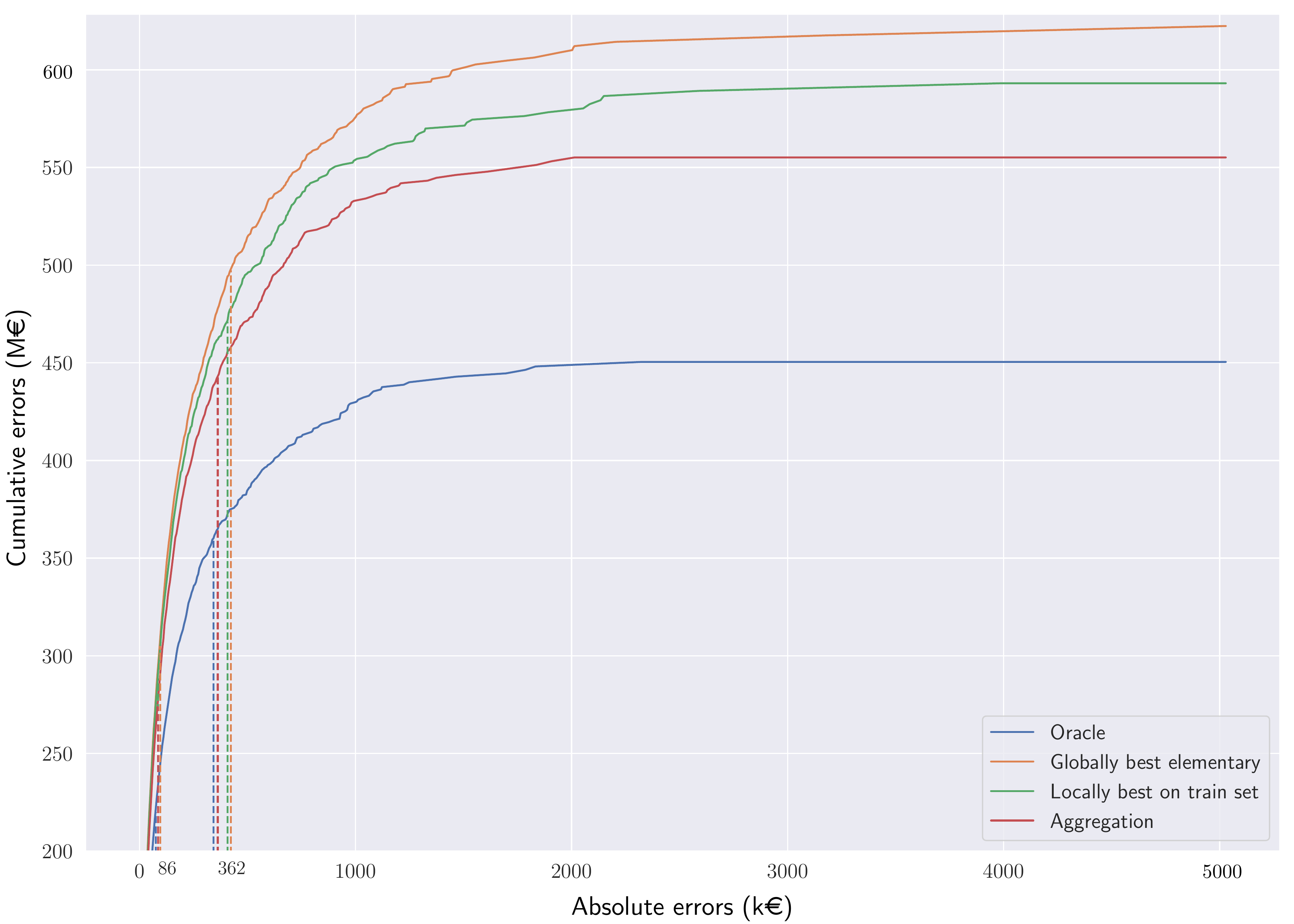} \vspace{-.5cm}
\end{center}
\caption{\small \label{fig:rob:cum} Cumulative absolute errors ($y$--axis, units: M€)
according to absolute errors ($x$--axis, units: k€), for three meta-predictors
(the locally best predictors on the train set, the globally best predictor on the test set,
and the oracle, which corresponds to the locally best predictors on the train set)
and an aggregation algorithm (ML-Poly with the absolute loss, no gradient trick, no projection),
for the case $(h,n)=(7,1)$, i.e., for 6-week-ahead-forecasting of 1 week of sales.
The dotted vertical lines indicate for each curve the preimages of $50\%$ and $80\%$
of the total cumulative errors; e.g., for the oracle, $50\%$ (respectively, $80$\%) of the total error
is suffered with individual errors smaller than 86 k€ (respectively, 362 k€).}
\end{figure}

Figure~\ref{fig:rob:cum} explains where differences in performance
between the locally best predictors on the train set, the globally best
predictor on the test set, and the aggregation algorithm lie: not
on small absolute errors (less than 90 k€, say), but
half on medium-sized errors (between 90 and 700 k€, say) and
half on large errors (more than 700 k€, say).

\begin{figure}[p]
\begin{center}
\includegraphics[width=0.9\textwidth]{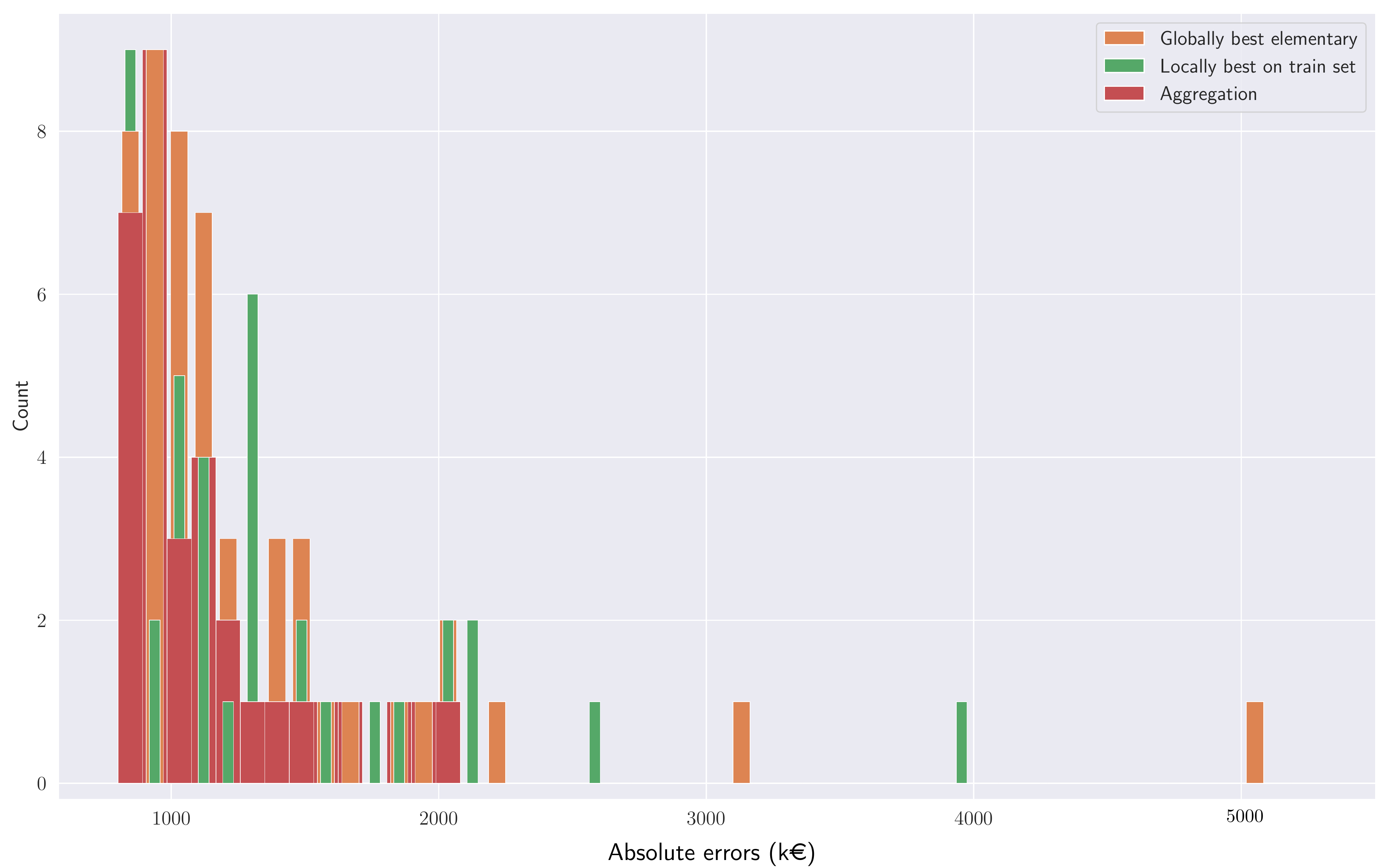} \vspace{-.4cm}
\end{center}
\caption{\label{fig:rob:large} \small Histogram count of large absolute errors (larger than 600 k€, see $x$--axis)
for two meta-predictors (the locally best predictors on the train set and the globally best predictor on the test set)
and an aggregation algorithm (ML-Poly with the absolute loss, no gradient trick, no projection),
for the case $(h,n)=(7,1)$, i.e., for 6-week-ahead-forecasting of 1 week of sales.}
\end{figure}

Figure~\ref{fig:rob:large} shows that there are not many
errors that are larger then 700 k€ out of the 156,208
errors considered: fewer than 40 or so. Yet, they account for a significant
part of the difference in performance. The aggregation algorithm considered
(still ML-Poly with the absolute loss, no gradient trick, no projection)
gets fewer of these large errors, and the maximal error it suffers equals
about 2 M€, while the maximal error for
the locally best predictors on the train set and for the globally best
predictor on the test set equal about 4 M€ and 5 M€, respectively.

\begin{figure}[p]
\begin{center}
\includegraphics[width=0.9\textwidth]{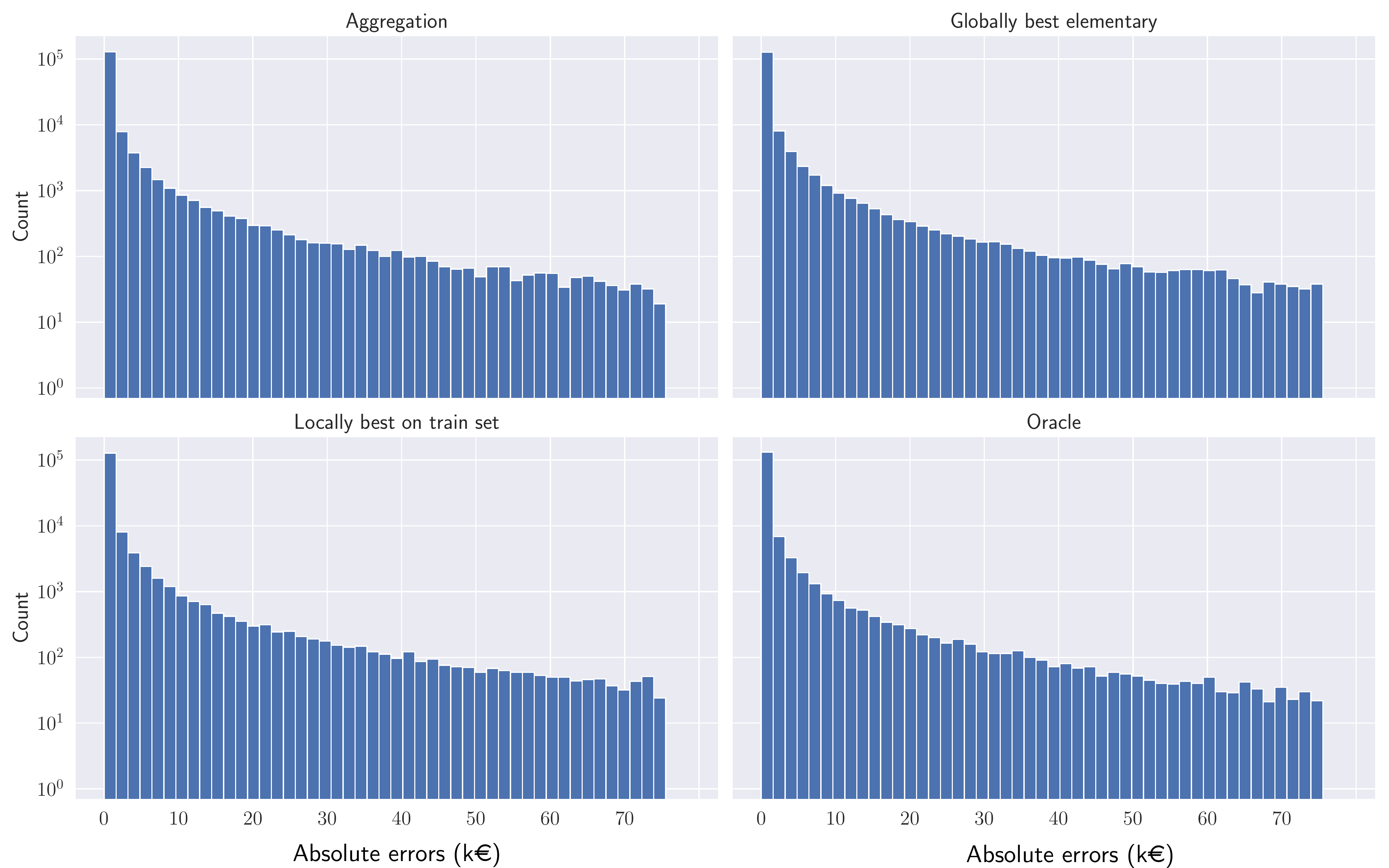} \vspace{-.4cm}
\end{center}
\caption{\label{fig:rob:small} \small Histogram counts of small absolute errors (smaller than 75 k€, see $x$--axis)
for three meta-predictors
(the locally best predictors on the train set, the globally best predictor on the test set,
and the oracle, which corresponds to the locally best predictors on the train set)
and an aggregation algorithm (ML-Poly with the absolute loss, no gradient trick, no projection),
for the case $(h,n)=(7,1)$, i.e., for 6-week-ahead-forecasting of 1 week of sales.}
\end{figure}

Figure~\ref{fig:rob:small} depicts the histograms of the small absolute
errors (smaller than 75 k€). These histograms are, first, virtually indistinguishable,
and second, account for most of the errors: they contain
almost all of the 156,208 absolute errors considered.
Yet, this is not where differences in performance mostly take place.

\vfill

\clearpage

\clearpage
\subsection{Evolution of the Weights Issued by the Aggregation Algorithms}
\label{sec:weights}

The aggregation algorithms considered in Section~\ref{sec:aggreg} issue convex weights:
at each prediction step, the forecast $\wh{y}^{(j)}_{t+h,\gamma}$ of the $j$--th elementary predictor is assigned a weight
$w^{(j)}_{t+h,\gamma}$ and an aggregated forecast is formed according to
\[
\sum_{j=1}^J w^{(j)}_{t+h,\gamma} \, \wh{y}^{(j)}_{t+h,\gamma}\,.
\]
The vectors $\underline{w}_{t+h,\gamma} = \bigl( w^{(j)}_{t+h,\gamma} \bigr)_{1 \leq j \leq J}$ are convex weight vectors:
their elements are nonnegative and sum up to~$1$.
A natural question is: do they have any particular structure? Do they converge, e.g., to a Dirac mass on
a given elementary predictor?

\begin{figure}[p]
\vspace{-.5cm}
\begin{center}
\begin{tabular}{cc}
\multicolumn{2}{c}{\includegraphics[width=0.38\textwidth]{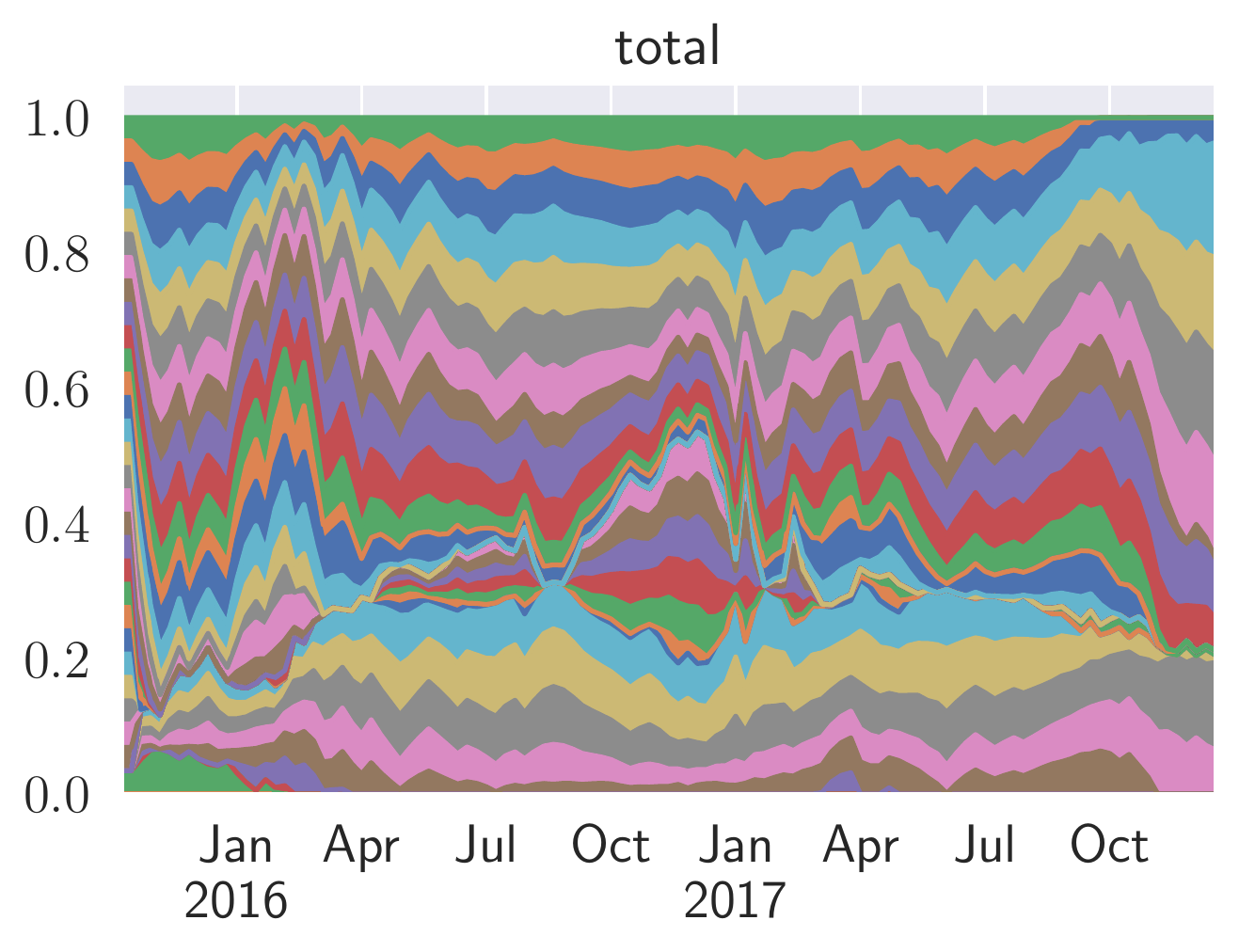}} \\
\hline
\ & \ \\
\includegraphics[width=0.38\textwidth]{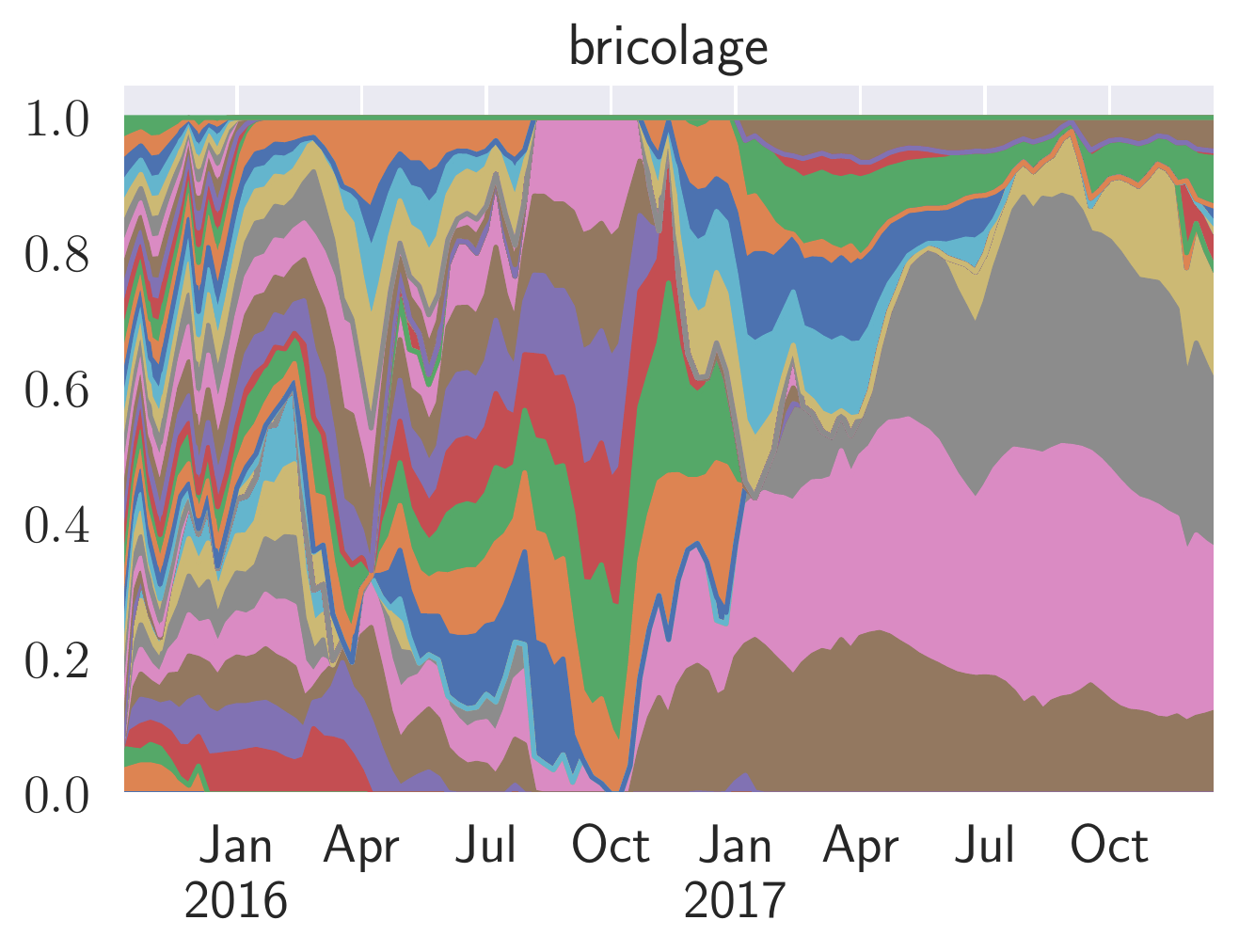}
& \includegraphics[width=0.38\textwidth]{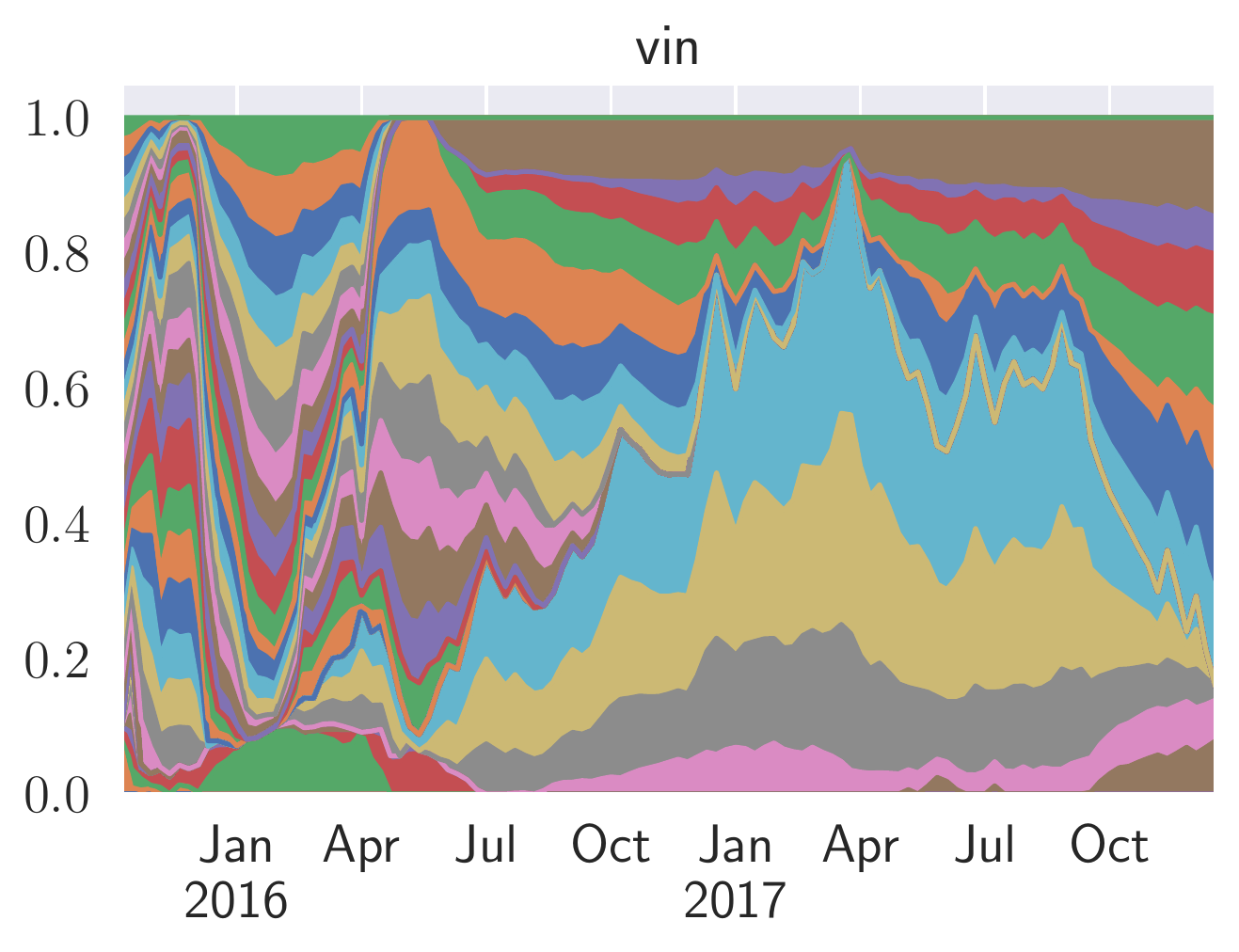} \\
\includegraphics[width=0.38\textwidth]{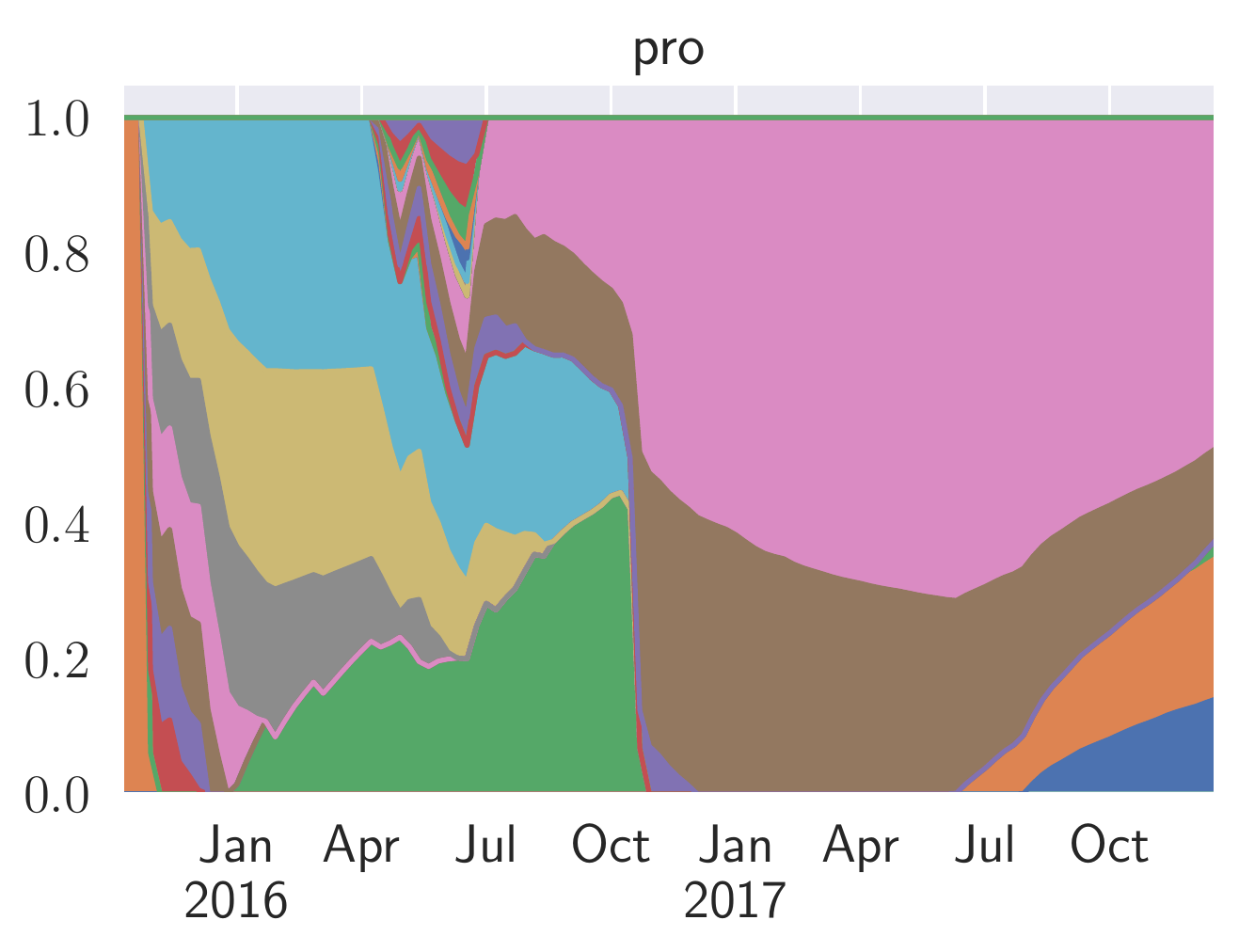}
& \includegraphics[width=0.38\textwidth]{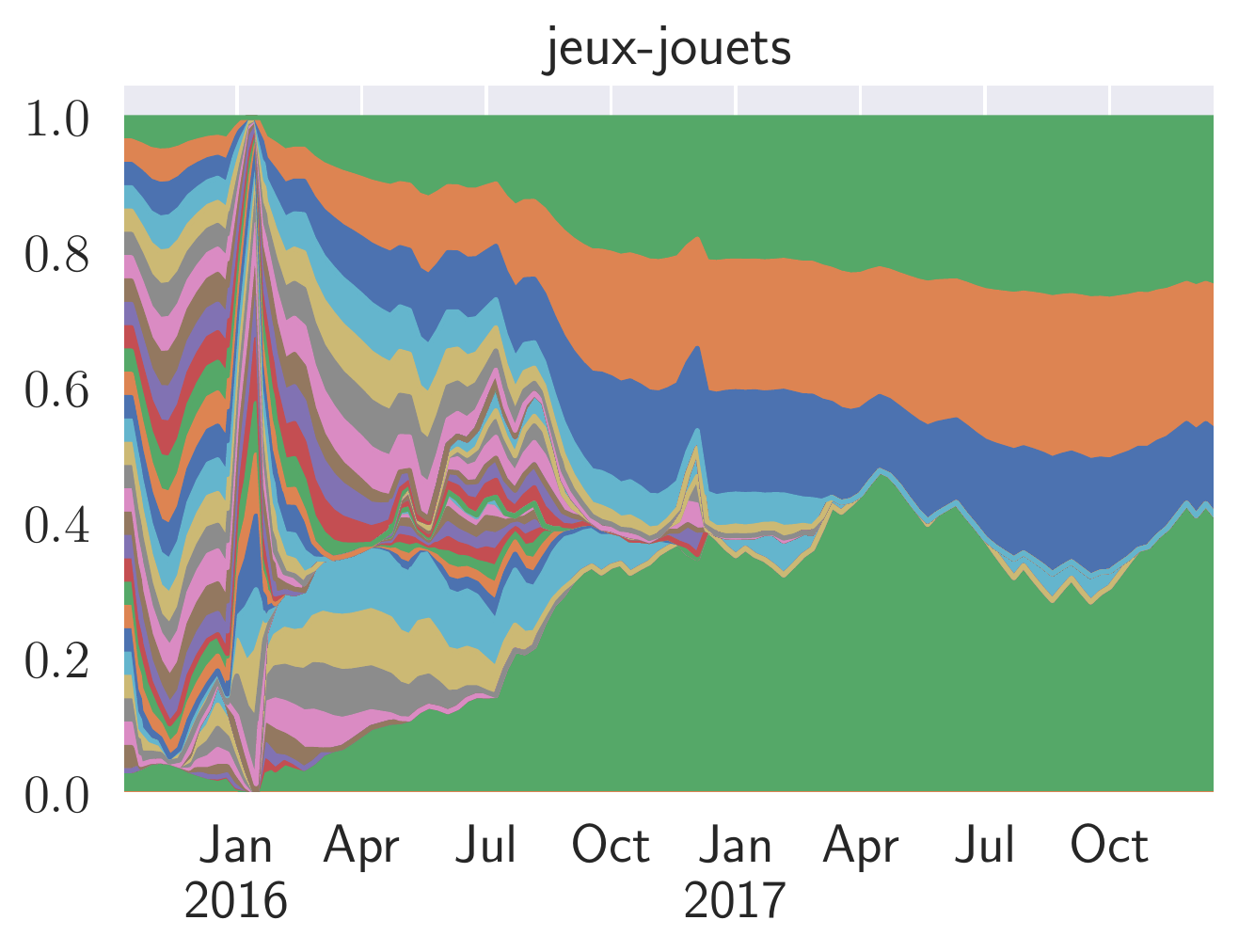} \\
\includegraphics[width=0.38\textwidth]{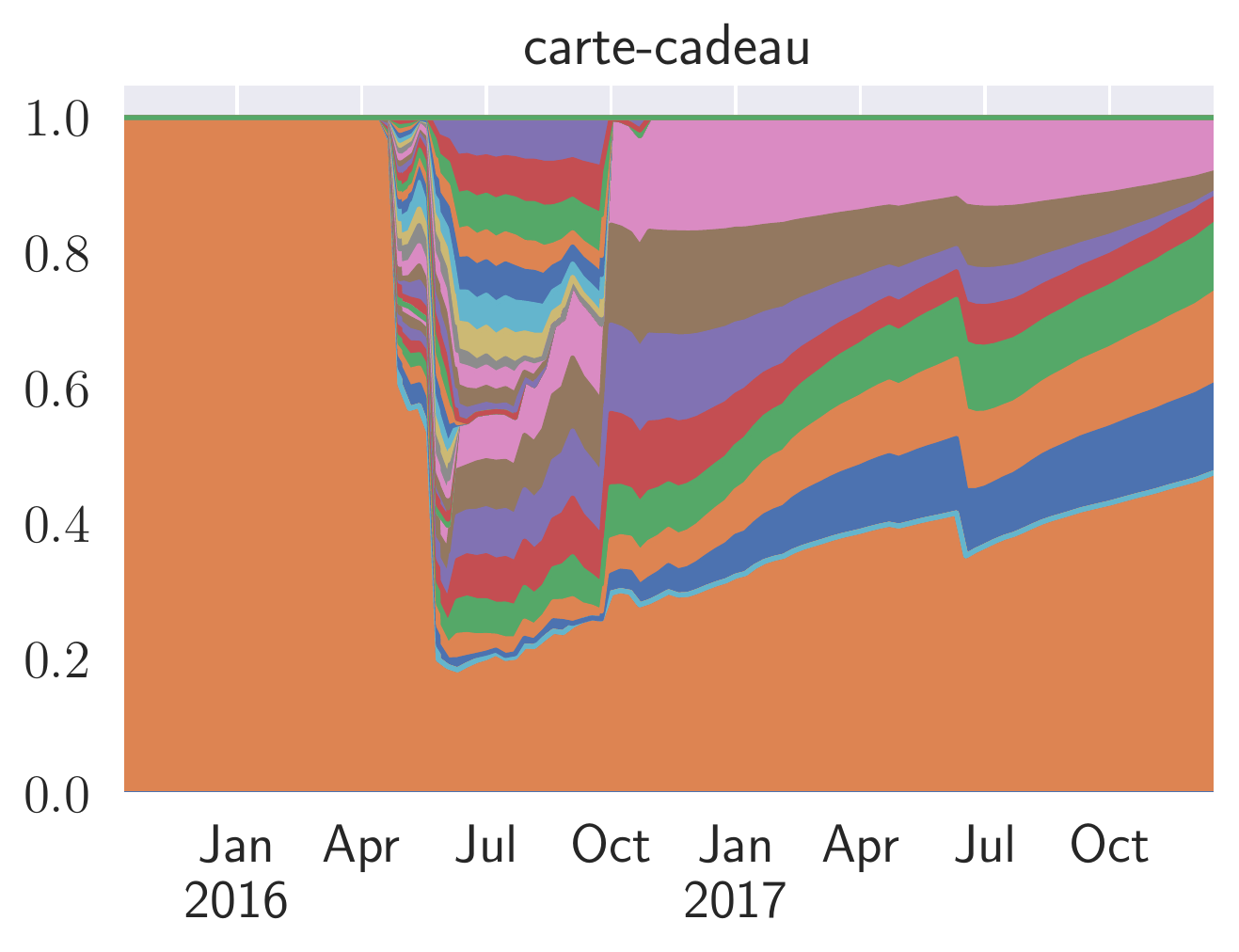} &
\includegraphics[width=0.38\textwidth]{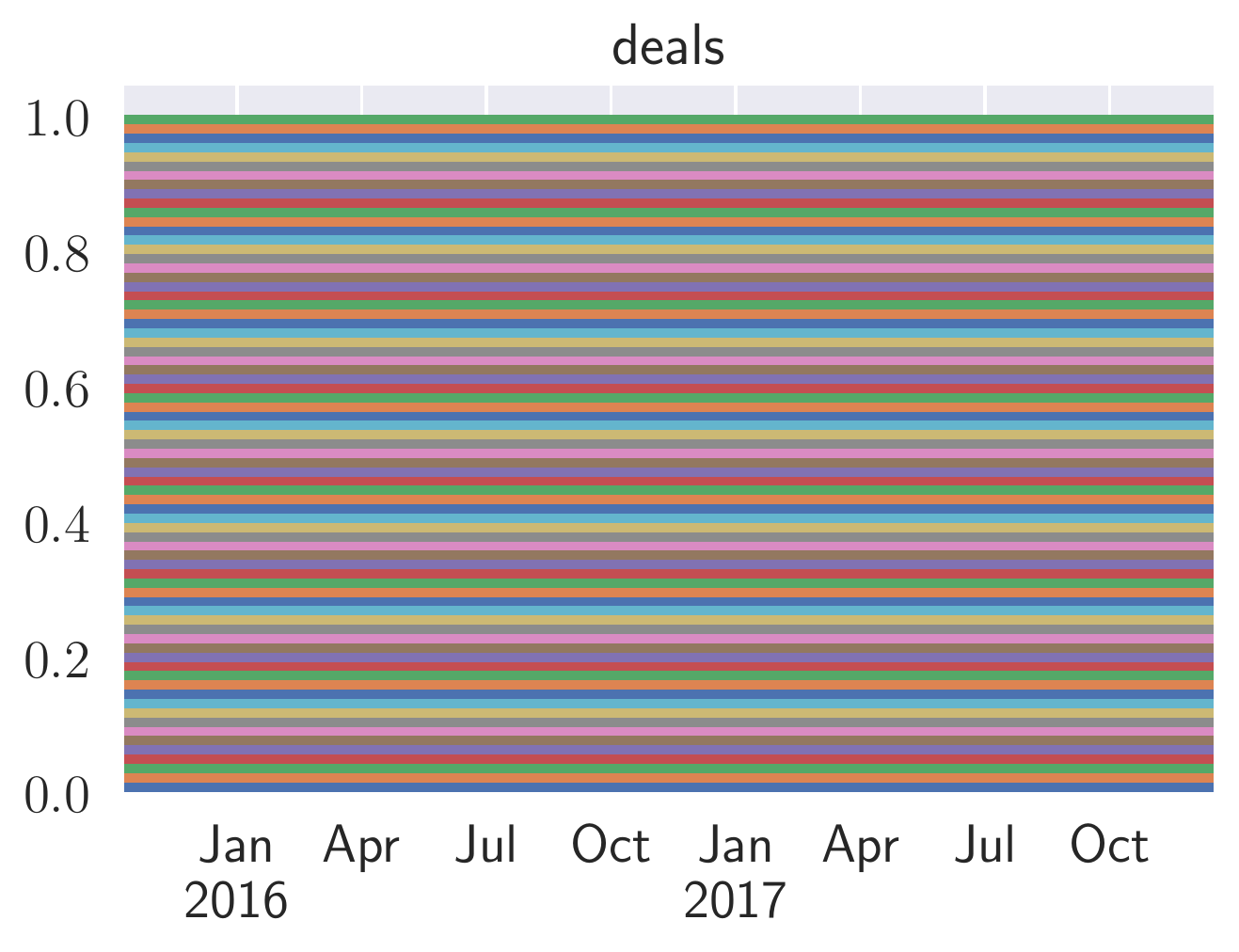}
\end{tabular}
\vspace{-.5cm}
\end{center}
\caption{\label{fig:weights} Evolution of the convex weights put on each elementary predictor
over time (from the point in time when all elementary predictors are defined: end of year 2015
to end of year 2017) by ML-Poly (with the absolute loss, no gradient trick, and no projection step),
for the prediction of total sales (first line) and the sales of
a representative subset of families (lines 2, 3 and 4); namely, from left to right and
from top to bottom: DIY--supplies (``bricolage''), wine (``vin''), equipment for professionals (``pro''),
toys (``jeux--jouets''), gift cards (`carte-cadeau''), special offers (``deals'',
for which the entire series of sales is null).
The weights for each of the $73$ elementary predictors are associated with a given color
on a given graph and sum up to $1$.
}
\end{figure}

Section~\ref{sec:holt} defined $J=73$ elementary predictors.
Figure~\ref{fig:weights} depicts the evolutions of the weight vectors picked over time
ML-Poly (with the absolute loss, no gradient trick, and no projection step)
for the root note (the total sales) and 6 families, which form a representative subset
of the 53 families. The main observation is that weights never converge to a Dirac mass
on a given elementary predictor. For all cases depicted, at least 5 or 6 elementary predictors,
and typically rather 10--15 of them, are used. We see that weights evolve significantly over time,
sometimes in a smooth way, sometimes in a more radical way.
Only one picture depicts no evolution at all (weights remain uniform): it corresponds to the family ``deals'',
which is one of the 6 families for which
the entire series of sales are null (see Table~\ref{tab:descr}).

The evolutions depicted on Figure~\ref{fig:weights} illustrate that aggregation algorithms
are reactive to changes and may reallocate the weights put on elementary predictors when needed.
This is in contrast with a meta-predictor like the locally best predictors on the train set,
which would need to be recomputed periodically from scratch to accommodate changes.

\vspace{2cm}
\subsection*{Acknowledgements}
We thank Ludovic Schwartz for his help on preliminary results that led to this article,
during his internship in Spring 2018.
This work was financially supported by AMIES [``Agence pour les math{\'e}matiques en interaction avec l'entreprise et la soci{\'e}t{\'e}'',
a French agency dedicated to interactions of mathematics with business and society] and the company Cdiscount.

\clearpage
\bibliographystyle{plainnat}
\bibliography{Agregation-Holt-Cdiscount}

\end{document}